\documentclass[journal,comsoc]{IEEEtran}
                                                          
\usepackage[dvipsnames]{xcolor}
\usepackage{threeparttable}
\usepackage{makecell}
\usepackage{pifont}
\newcommand*{\na}[1]{\textcolor{black}{#1}}
\newcommand*{\ik}[1]{\textcolor{black}{#1}}

\newcommand{\cmark}{\ding{51}}%
\newcommand{\xmark}{\ding{55}}%
\IEEEoverridecommandlockouts                              
\usepackage{graphicx}
\usepackage{multirow}
\usepackage{authblk}
\usepackage{url}

\title{\LARGE \bf
\ik{De-anonymisation attacks on Tor: A Survey}
 }

\author[1,2,*]{Ishan Karunanayake}
\author[1,2]{Nadeem Ahmed}
\author[1,2]{Robert Malaney}
\author[1,3]{Rafiqul Islam}
\author[1,2]{Sanjay Jha}
\affil[1]{Cyber Security Cooperative Research Centre (CSCRC) - Australia}
\affil[2]{University of New South Wales (UNSW) - Sydney, Australia}
\affil[3]{Charles Sturt University - Albury, Australia}
\affil[*]{Corresponding author - \textit{ishan.karunanayake@unsw.edu.au}}

\begin{document}

\maketitle
\thispagestyle{empty}
\pagestyle{empty}
\urlstyle{same}

\begin{abstract}

Anonymity networks are becoming increasingly popular in today's online world as more users attempt to safeguard their online privacy. Tor is currently the most popular anonymity network in use and provides anonymity to both users and services (hidden services). However, the anonymity provided by Tor is also being misused in various ways. Hosting illegal sites for selling drugs, hosting command and control servers for botnets, and distributing censored content are but a few such examples. As a result, various parties, including governments and law enforcement agencies, are interested in attacks that assist in de-anonymising the Tor network, disrupting its operations, and bypassing its censorship circumvention mechanisms. \ik{In this survey paper, we review known Tor attacks and identify current techniques for the de-anonymisation of Tor users and hidden services. We discuss these techniques and analyse the practicality of their execution
method. We conclude by discussing improvements to the Tor framework that help prevent the surveyed de-anonymisation attacks.}

\end{abstract}

\section{INTRODUCTION}

Over the past few decades, many online services have impacted the daily lives of Internet users. With that, a genuine concern has emerged as to how to browse the Internet while maintaining privacy. 
Privacy-preserving mechanisms over the Internet are all the more important for whistle-blowers and citizens of totalitarian governments, who are usually in dire need to protect their online identity. Other use cases of anonymous networks include sensitive communications of military and business organisations over the public Internet \cite{JEFFRIES2014}. The above reasons have led to the research and development of anonymous communication systems \cite{EDMAN2009}. The early anonymity systems such as Mix-Net \cite{CHAUM1981}, Babel \cite{GULCU1996}, and Mixminion \cite{DANEZIS2003MIXMINION} were not widely adopted as they suffered from high latency issues and are now superseded by low-latency systems, as we now discuss. 

The Onion Router project, which is more commonly known as Tor \cite{DINGLEDINE2004}, is the most popular low latency anonymity network to date. Tor provides anonymity to users and supports the deployment of anonymous services, known as hidden services. However, as the anonymity provided by Tor was available to everyone, it quickly became an accessory to cybercrime and other criminal activities \cite{CHRISTIN2013}, \ik{as well as a tool for terrorists to anonymously spread their propaganda \cite{WEIMANN2016}.} This forced Law Enforcement Agencies (LEA) and governments to find ways to break its anonymity. 

Tor being an anonymity network, the most common objective of a Tor attack is to de-anonymise its users and services through de-anonymisation attacks. \na{In response to de-anonymisation, pro-anonymity researchers attempted to strengthen users' expected anonymity by improving security and fixing known bugs. The Tor network also grew in size over the years. This growth, combined with the security improvements, played a vital role in securing the Tor network against most legacy de-anonymisation attacks.}



\ik{In this paper, we conduct an updated survey with a comprehensive overview of de-anonymisation attacks on the Tor network. Although there are some past surveys of Tor attacks, most consider all types of Tor attacks, while we have narrowed the scope to only include de-anonymisation attacks. Our survey covers about 30 more de-anonymisation attacks than most past surveys \cite{SALO2010, ALSABAH2016,SALEH2018, CAMBIASO2019}. Moreover, we discuss more than 15 de-anonymisation attacks published after 2016 (not reported in most of the previous works \cite{ALSABAH2016,EVERS2016,ERDIN2015}), including attacks that use advanced techniques such as deep learning. 
We also note that some survey works \cite{ ERDIN2015,NEPAL2015,BASYONI2020} do not include details on website fingerprinting attacks and hidden service attacks, which are important types of de-anonymisation attacks.}

\ik{It is useful to have a well-defined taxonomy focusing on de-anonymisation attacks. In this regard, we present a new multi-level taxonomy to categorise de-anonymisation attacks on Tor. We use different discriminating factors at each level of the taxonomy to deliver a systematic categorisation. We classify and discuss each attack, focusing on the Tor circuit component(s) used and the method of execution. We also try to draw conclusions on the practicality of those attacks. Finally, we provide insights into how Tor's research has impacted its development since its initial deployment. 
We highlight several significant milestones over the years that are relevant to de-anonymisation attacks and discuss how security improvements have made some of the previously possible attacks unfeasible. We hope that this work will be a valuable resource for anyone who wants to gain knowledge or engage in research in this area.}


 

The structure of the paper is as follows. Section II provides the necessary background information required to understand the content of this paper, while Section III compares the scope of our survey with other related work. Section IV introduces and explains our proposed taxonomy. Details of individual de-anonymisation attacks are presented in Section V, mainly focusing on their method of execution. 
In Section VI, we discuss the evolution of the Tor network and its security, highlighting the applicability of existing attacks on the live Tor network. 
We conclude in Section VII with a discussion on potential directions for future work.

\section{BACKGROUND}

\begin{figure}[h]
\centering
\includegraphics[width=8cm]{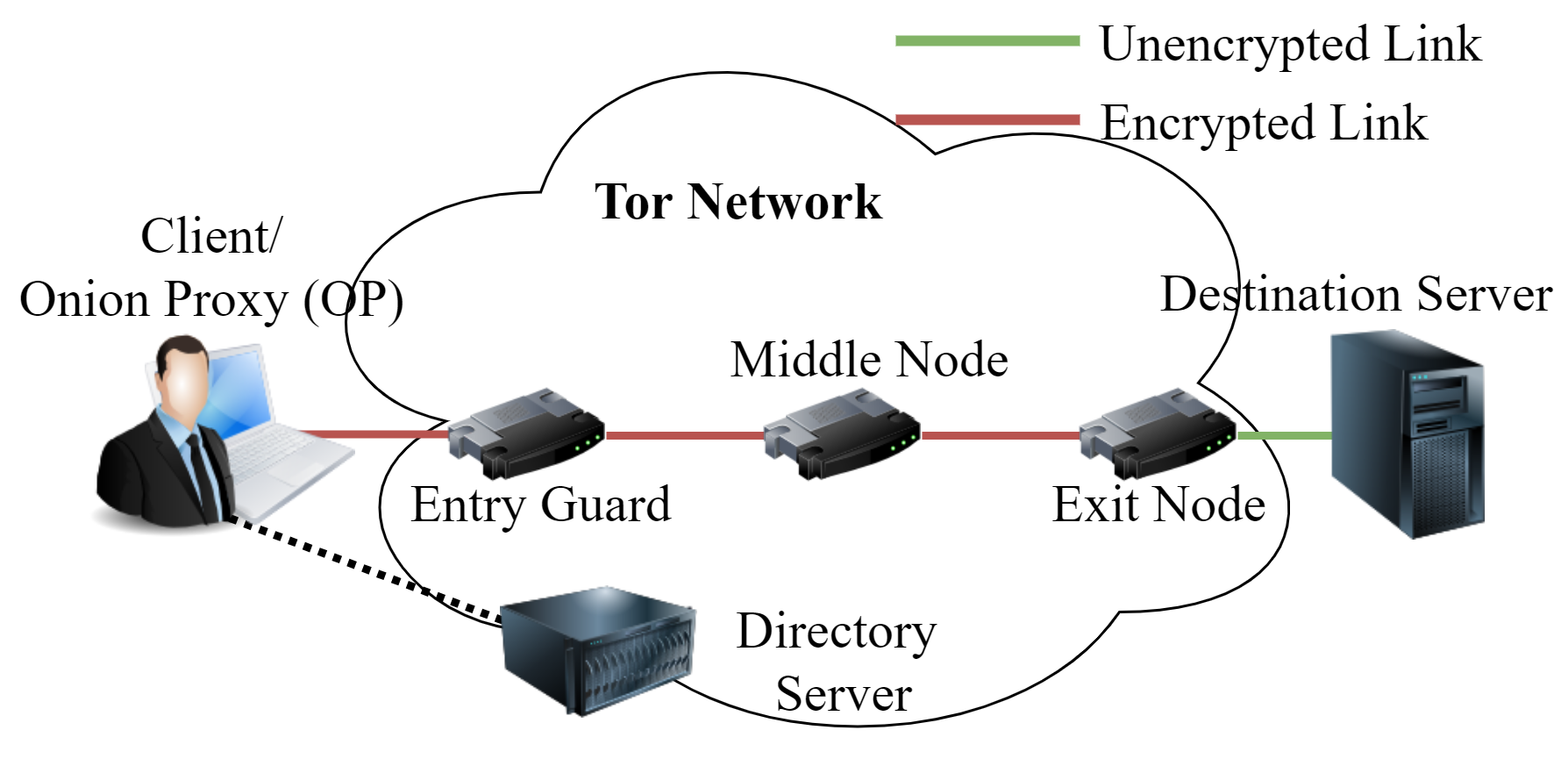}
\caption{Components of the Tor Network (standard Tor circuit)}
\label{fig:Tor Network}
\end{figure}

\begin{figure}[h]
\centering
\includegraphics[width=8cm]{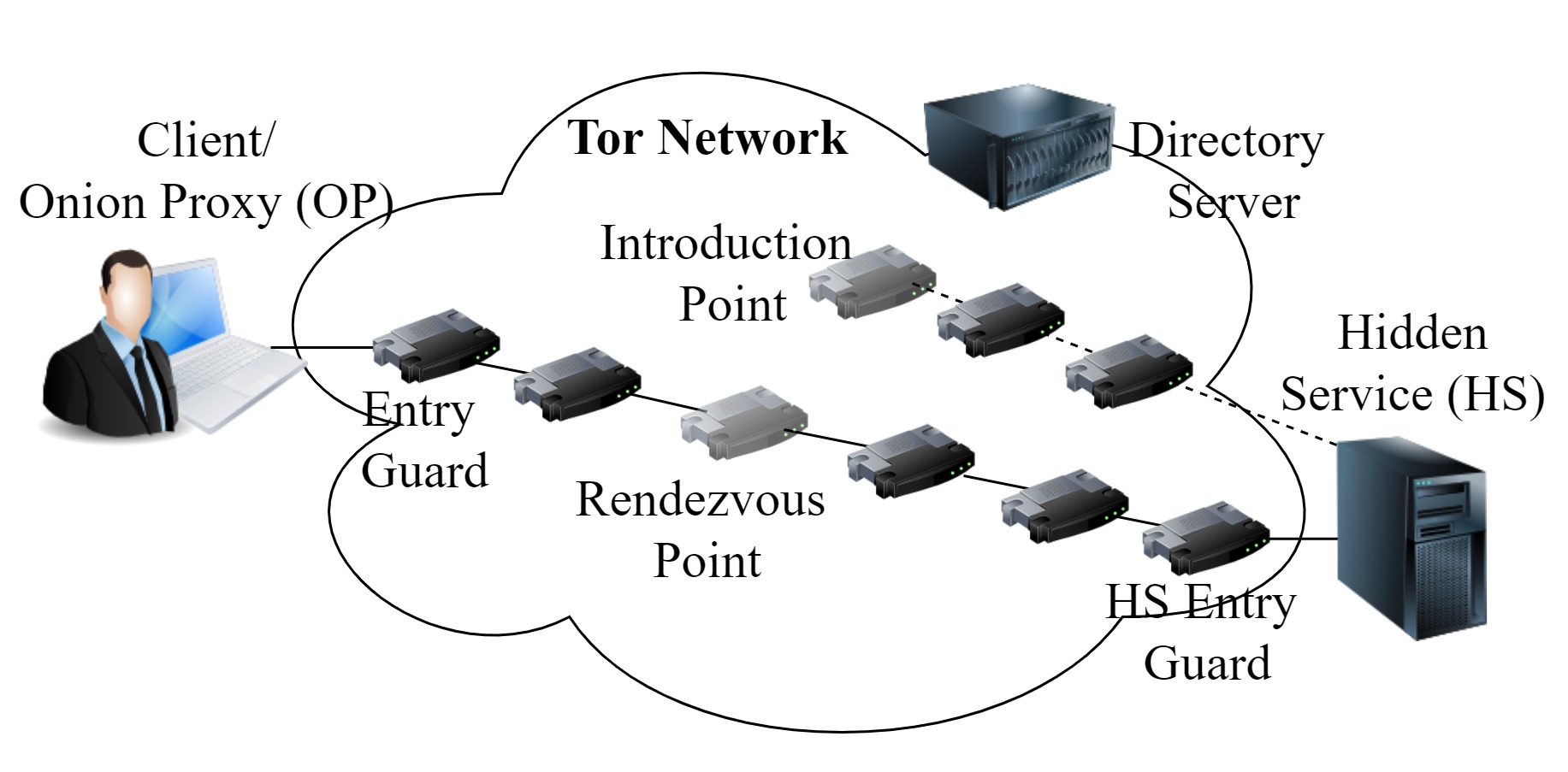}
\caption{Components of a Hidden Service}
\label{fig:Hidden services}
\end{figure}

In this section, we provide some background information to explain our taxonomy and the attacks discussed in this paper. The Tor network \cite{DINGLEDINE2004}, which is one of the most widely used anonymity networks today (along with other popular networks such as I2P \cite{ZANTOUT2011} and Freenet \cite{CLARKE2001}), has been using the concept of onion routing \cite{GOLDSCHLAG1999}. Tor is an overlay network based on Transmission Control Protocol (TCP) that builds circuits from a user to the destination server, which generally consists of three voluntary relays.\footnote{We use the terms nodes, routers and relays interchangeably throughout this paper.} Figures \ref{fig:Tor Network} and \ref{fig:Hidden services} show the components of a Tor network for a standard circuit, and hidden services respectively. The descriptions of some of the key components and their features are as follows.

\begin{figure*}[h]
\centering
\includegraphics[width= \textwidth]{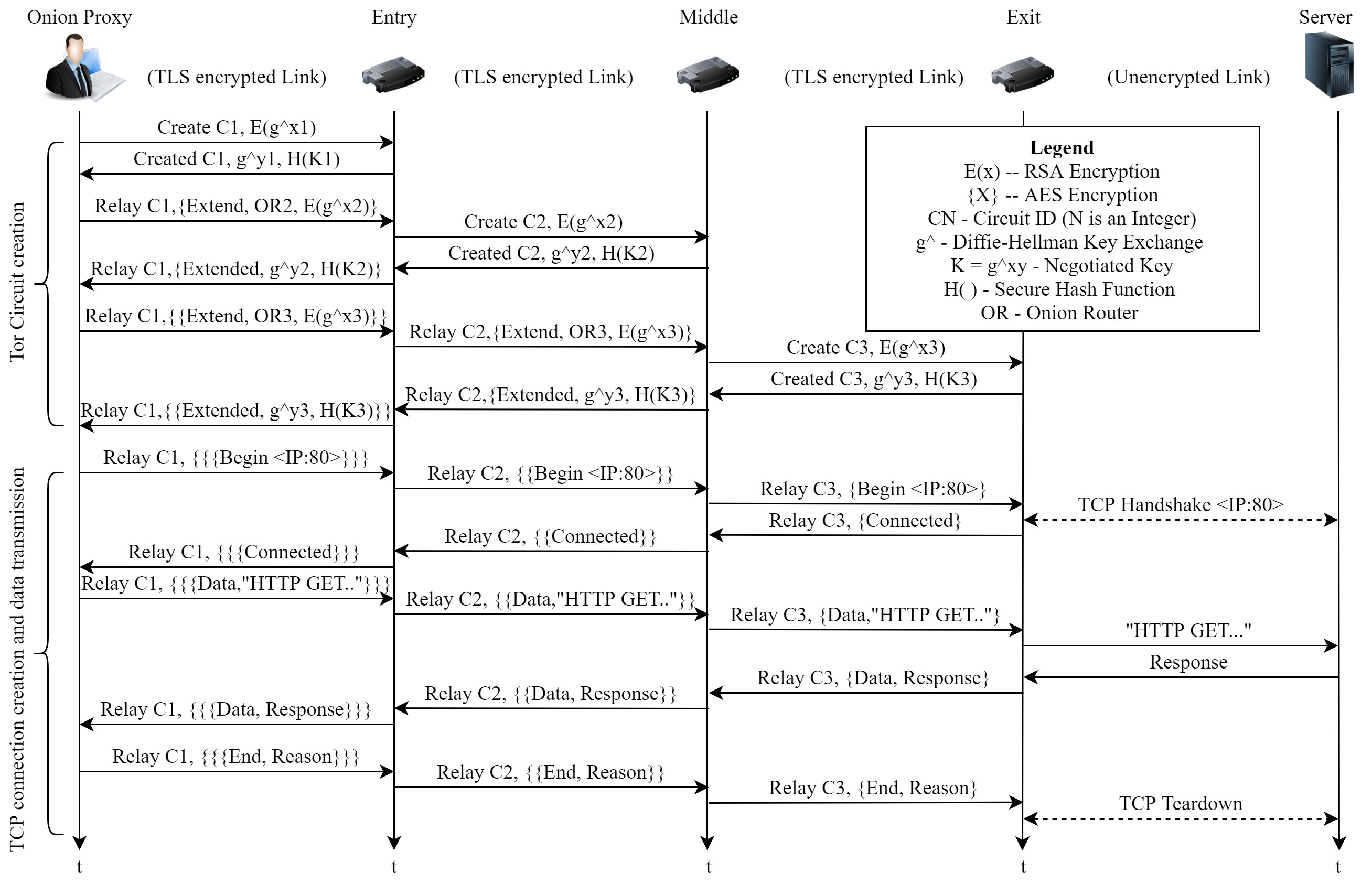}
\caption{Tor circuit creation and data transmission}
\label{fig:Tor circuit creation}
\end{figure*}

\begin{itemize}
    \item \textbf{\textit{Onion Proxy (OP)}}: This is a small piece of local software that needs to be installed on the user's device. It enables communication with the directory servers (DSs), establishes connections in the Tor network, and handles connections from the user's applications. In this paper, we also refer to this as the Tor client.
    \item \textbf{\textit{Directory Servers (DS)}}: These are a small set of trusted and known servers in the network that actively keep details about the status of the complete network. DSs produce a consensus document with the status of the network relays, bandwidth availability, exit policies, etc. The OPs can download this document from a DS and select three suitable relays to establish the communication circuit to a destination.
    \item \textbf{\textit{Entry Node/Guard}}: This is the relay in the Tor network that the client directly connects to, and hence, it knows the Internet Protocol (IP) address of the client. Therefore, several early Tor attacks either compromised existing entry nodes or installed new nodes to participate as entry nodes to de-anonymise users. We discuss these de-anonymisation attacks in Section V. Another important feature of entry nodes came with the introduction of guard nodes. As Tor creates new circuits quite frequently, there was always a chance that at some point, it would select an adversary-controlled node as the entry node. Tor network introduced \textit{Guard nodes} to reduce the probability of this occurrence. Now, OPs select a small set of trusted nodes as guard nodes and use only one of these nodes as the entry node for all circuits until they (OPs) pick a different set of nodes as guards. DSs assign a \textit{Guard Flag} to a node after considering its bandwidth, uptime and time in the Tor network. Any node is eligible to become a guard node on the eighth day after joining the Tor network \cite{NODELIFECYCLE}.
    \item \textbf{\textit{Exit Node}}: This is the final hop of the Tor circuit. Therefore, it knows the IP address of the destination server accessed via the Tor network. Moreover, as the last layer of encryption provided by the Tor network ends here (unless the client's application is also using end-to-end encryption such as TLS), a malicious exit node can easily observe the Tor traffic flowing through it.
    \item \textbf{\textit{Hidden Services (HS)}}: \na{By default, Tor provides anonymity to the user but does not hide the identity of the website that the user is accessing. An entity with access to traffic at the exit node of the Tor circuit or the link between the exit node and the website can retrieve the website's IP address. The Tor network supports Hidden Services (HS), also known as Onion Services, to address this issue. HS can be hosted on a node inside the Tor network or an external node.} These have a top-level domain name ending in \textit{.onion}. The HS owner can advertise this onion address over the public Internet. A potential client has to find out about this service address from the web or other similar means. The anonymity provided by HS attracts those engaged in criminal and unethical conduct, including those who sell drugs \cite{CHRISTIN2013} and child pornography \cite{GUITTON2013}, forcing LEAs to identify and shut down these services.
    \item \textbf{\textit{Introduction Points}}: These are random nodes selected by the HS to register its services with the Tor network. To avoid any impact from possible Denial of Service (DOS) attacks against a single introduction point, the HS usually selects several of them. The HS then advertises these selected introduction points and its public key in the Hidden Service Directories (HSDirs)\footnote{HSDirs are a type of DSs with some specific properties, which are used to publish the service descriptors of HSs. We use DS and HSDir interchangeably when referring to HS circuit creation throughout the paper.}. The introduction points do not know the IP address of the HS as they are connected to the HS via a complete Tor circuit consisting of multiple intermediate relays.
    \item \textbf{\textit{Rendezvous Points (RPs)}}: This is a random Tor node selected by the client OP before the client initialises a connection with any of the introduction points advertised by the DS. \ik{The client selects two other nodes (entry and middle) and establishes a Tor circuit to the RP via these nodes. As a result, the RP does not know the identity of the client.}
    \item \textbf{\textit{Bridges}}: As DSs maintain a list of relays in the Tor network to advertise to all clients, this information could easily be used by a service provider to censor and block the Tor network. To mitigate this issue, bridges were introduced. Bridges are normal Tor relays that are not listed publicly in the main Tor directory. They replace guard nodes in the circuit; however, only a few bridges are provided to each client. Therefore, no authority is able to obtain a complete list of bridge nodes. It is not necessary to have bridges as middle or exit relays as the bridge enables encrypted connections to censored Tor relays. In addition, having bridges as middle and exit relays would require more bridges to be published for a single client, rendering them useless. There are a few ways in which users can obtain these bridge addresses. They can visit the Tor project website, email the Tor project team or request bridges through the Tor browser. 
\end{itemize}



Having explained different components of the Tor network, we will now discuss how a typical Tor circuit is established. 
\subsection{\ik{Standard Tor circuit establishment}}
Before communicating over the Tor network, a Tor client must establish a circuit through the Tor network. The user is required to have the Onion Proxy (OP) installed on the device being used for browsing. The OP first contacts a DS and requests a list of active relays in the network. Then it selects three relays from the list to act as the entry, middle, and exit nodes, and incrementally creates a circuit by exchanging encryption keys with each node, one hop at a time \cite{DINGLEDINE2004}. The key exchange is done via the Diffie-Hellman handshake \cite{LI2010}, as shown in Figure \ref{fig:Tor circuit creation}. Once this connection consisting of three hops has been established (Figure \ref{fig:Tor Network}), the user can now communicate with the intended destination server over the established circuit.

\begin{figure}[h]
\centering
\includegraphics[width=8cm]{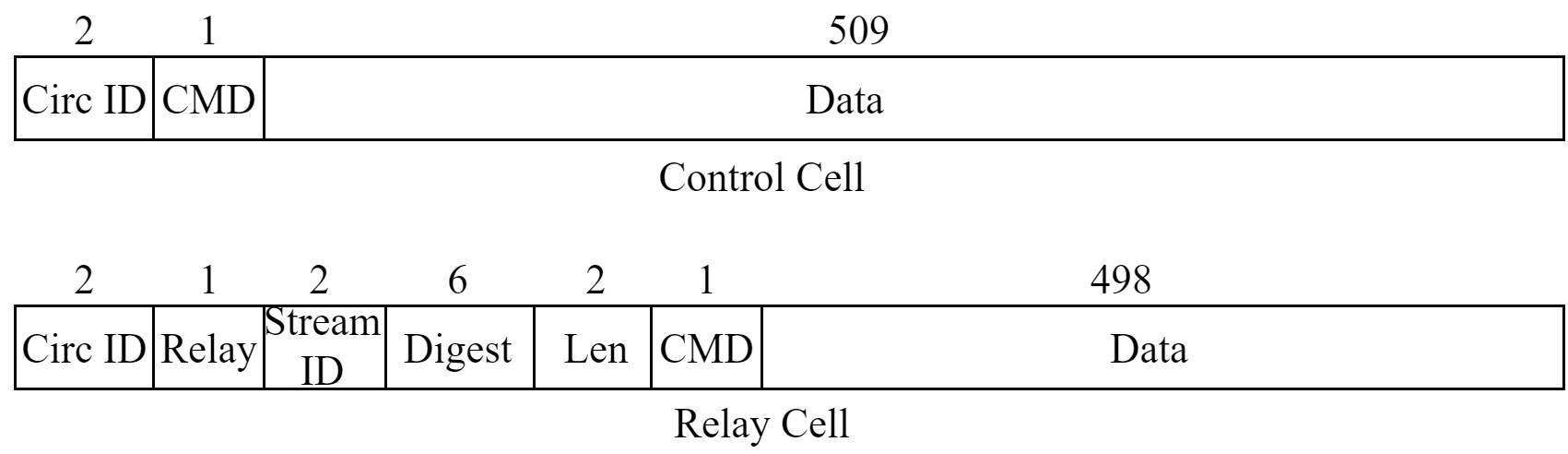}
\caption{Tor cells}
\label{fig:Tor Cell}
\end{figure}

\begin{figure*}[h]
\centering
\includegraphics[width=14cm]{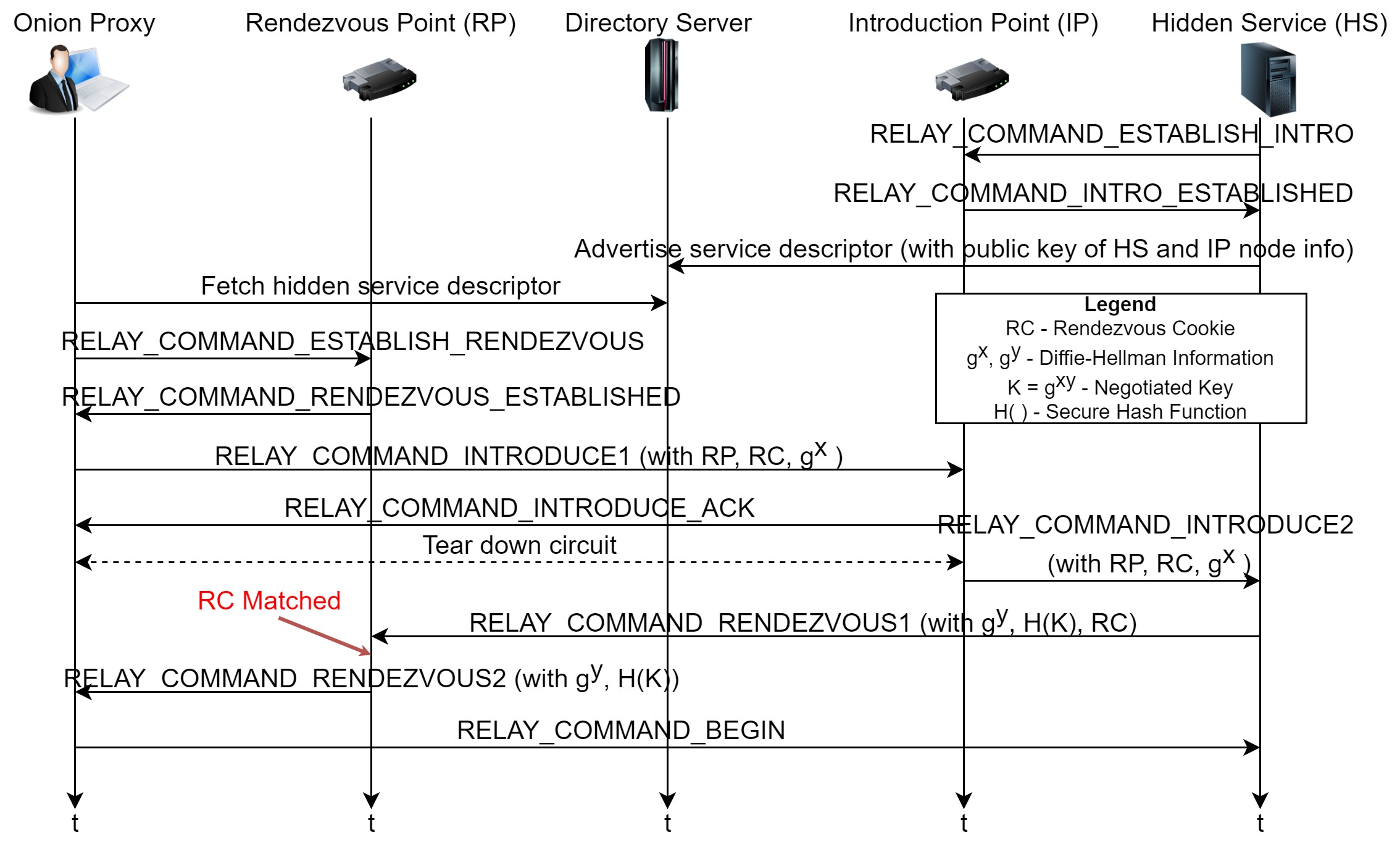}
\caption{Hidden server connection establishment}
\label{fig:Hidden server connection}
\end{figure*}

Tor uses fixed-length cells of 512 bytes for communication to make traffic analysis harder \cite{DINGLEDINE2004}. There are two types of cells; control cells and relay cells. Figure \ref{fig:Tor Cell} shows the structure of these two cell types. Control cells are always interpreted at the receiving nodes and they issue commands such as \textit{create, created, destroy} or \textit{padding}. Relay cells carry end-to-end data and consist of an additional relay header. This relay header includes a stream ID (as multiple streams are multiplexed over a single circuit), an end-to-end checksum for integrity checking, a payload length and a relay command. The relay command can be \textit{relay data, relay begin, relay end, relay teardown, relay connected, relay extend, relay extended, relay truncate, relay truncated,} or \textit{relay sendme}. The relay header and the payload are encrypted with the 128-bit counter mode Advanced Encryption Standard (AES-CTR), which uses symmetric keys negotiated via the Diffie Hellman Key Exchange (DHKE) \cite{LI2010}.

\subsection{\ik{Circuit establishment for Tor HS}}
The establishment of a connection between a user and a HS is shown in Figure \ref{fig:Hidden server connection}. It should also be noted that there are multiple relays\footnote{\ik{For example, as shown in Figure \ref{fig:Hidden services}, the connections between the client and RP, and the HS and IP have two intermediate nodes (entry node - middle node). Meanwhile, the client connects to the IP, and the HS connects to the RP via three intermediate nodes. These intermediary nodes help protect the client's or HS's anonymity if the IPs or the RP are compromised.}} involved between the components shown in Figure \ref{fig:Hidden server connection} although we have only displayed the sending and receiving ends of the relevant communications. Firstly, the HS selects multiple introduction points from the available nodes in the Tor network and builds connections to those nodes. Following this, it connects to the DS and advertises a \textit{service descriptor} with the HS's public key, expiration time, and the details of the selected introduction points. The HS owner can then advertise the service's onion address using multiple platforms (e.g. websites, blogs, other hidden services). If users want to access a HS, they need to find its onion address through these platforms. When the user searches an address in the browser, the OP fetches the service descriptor of that particular HS from the DS. This way, the OP finds out about the HS's introduction points and its public key. The OP then selects an RP, establishes a Tor circuit to the RP (via two nodes (entry and middle)), and sends a message with the RP's address, and a \ik{one-time secret} called the \textit{Rendezvous Cookie (RC)} to one of the introduction points. The introduction point forwards this message, which is encrypted with the HS's public key, to the HS. Once the HS receives the message, if it wants to establish a connection with that client, \na{it (HS) selects three Tor nodes (one entry and two middle) and creates a three-hop connection to the RP, which keeps the HS's identity anonymous from RP. Following this, the client and the HS can communicate using the six-hop circuit via the RP, as shown in Figure \ref{fig:Hidden services}} in the same way they communicate with a traditional web service \cite{LING2013}.

\ik{For a reader who wishes to learn more about how Tor works, we recommend reading the Tor design paper \cite{DINGLEDINE2004}. This paper will help the reader to understand all the basic concepts behind Tor, its components, circuit creation, threat model, design goal, assumptions, etc. However, after its initial deployment, there have been some important additions to Tor, such as guard nodes, bridges, Tor browser bundle, etc. We recommend following the three-part article series \cite{MATHEWSON2012_1,MATHEWSON2012_2,MURDOCH2012_3} in the Tor blog to obtain an idea about these developments. These resources provide the necessary background knowledge about Tor.}

\section{RELATED WORK}

In this section, we present other survey papers and related work that cover Tor attacks. In 2009, Edman \textit{et al.} \cite{EDMAN2009} published a survey on existing anonymous communication systems. Their survey mainly describes the research on designing, developing, and deploying such communication systems. Furthermore, the authors of \cite{EDMAN2009} discuss several adversarial models based on properties like capability, visibility, mobility, and participation. Their survey has a section on traffic analysis attacks categorised under website fingerprinting \ik{(WF)} \cite{HINTZ2002}, timing analysis \cite{LEVINE2004,MURDOCH2005}, predecessor attacks \cite{WRIGHT2004}, and disclosure attacks \cite{DANEZIS2003DISCLOSURE_ATTACK}. \cite{EDMAN2009} is a comprehensive survey of research on anonymous communication designs and approaches but not in terms of attacks.

In 2010, Salo \cite{SALO2010} attempted to survey and categorise Tor attacks into five categories:
1. probabilistic models that provide information about the Tor network based on mathematical modelling, 
2. attacks that attempt to compromise the victim's entry and exit nodes,
3. Autonomous System (AS) and global level attacks by a passive global adversary,
4. traffic and time analysis attacks, and
5. protocol vulnerabilities that address weaknesses in the Tor protocol.
However, Salo's work does not take into account attacks on HS \cite{OVERLIER2006, MURDOCH2006} as well as website fingerprinting attacks against Tor \cite{HERRMANN2009}, which began to emerge around the time \cite{SALO2010} was written.

A survey on de-anonymisation attacks against HS was conducted by Nepal \textit{et al.} \cite{NEPAL2015} in 2015, however, it is limited to the three attack schemes presented by Ling \textit{et al.} \cite{LING2013}, Jansen \textit{et al.} \cite{JANSEN2014}, and Biryukov \textit{et al.} \cite{BIRYUKOV2013}. Nepal \textit{et al.} explain the basic functionality of these attacks and provide a comparison between the attack schemes in terms of the simulation environment, the time required for de-anonymisation, the true positive rate, and the number of compromised nodes required to launch the attack successfully.

In 2015, Erdin \textit{et al.} published a survey paper on de-anonymisation attacks \cite{ERDIN2015}. However, in \cite{ERDIN2015}, the authors have focused only on the de-anonymisation of users. They discuss such type of attacks on both Tor and I2P \cite{ZANTOUT2011} networks. In \cite{ERDIN2015}, the authors explain de-anonymisation attacks under two categories: 1. Application-based attacks, and 2. Network-based attacks. Most of the time, Application-based attacks are a result of insecure applications or user's carelessness. Erdin \textit{et al.} discuss the attack vectors of Application-based attacks such as plugins, Domain Name System (DNS) lookups, java applets, active documents, and BitTorrent. In contrast, Network-based attacks exploit limitations or trade-offs of the anonymity network. The authors of \cite{ERDIN2015} discuss examples of network-based attacks under five approaches. 1. Intersection attacks, 2. Flow multiplication attacks, 3. Timing attacks, 4. Fingerprinting attacks, and 5. Congestion attacks. They explain how these attack types affect Tor and I2P networks and present potential remedies against these attacks. 

\ik{In a survey published in 2016, Alsabah \textit{et al.} \cite{ALSABAH2016} evaluate Tor research in several areas including performance and security. They use the following categories in their paper.}
1. Traffic management - Tor's congestion control, quality of service, etc. are discussed in this category under application layer and transport layer approaches.
2. Router selection - The chances of a Tor node being selected by the OP depend primarily on the node's bandwidth. However, there are other factors affecting this, including the node being a guard node. The research on Tor's router selection problem is explored in this category.
3. Scalability - Tor's scalability approaches are investigated here under a peer to peer approach and a scalable centralised approach that uses private information retrieval (PIR-TOR).
4. Circuit construction - Improving the computational overhead of Tor's circuit construction is discussed in this category.
5. Security - This section investigates Tor attacks and categorises them into active and passive attacks. Passive attacks are further categorised into AS level adversaries \cite{EDMAN2009AS} and website fingerprinting (this will be discussed in more detail in section V), while active attacks are sub-categorised into end-to-end confirmation attacks, path selection attacks, and side-channel information attacks. Alsabah \textit{et al.} discuss 22 attacks on the Tor network within the above categorisation.

\begin{table*}[t]
  \footnotesize
  \centering
  \caption{Summary of related work}
  \label{table:summary}
  \begin{threeparttable}
    \begin{tabular}{ |l|l|l|c|c|c|c|l| } 
      \hline
      \textbf{Publication} & \textbf{Year} & \textbf{Main Focus} & \thead{\textbf{\ik{No of \tnote{*}}} \\ \textbf{\ik{attacks} }} & \thead{\textbf{Include} \\ \textbf{WF attacks}} & \thead{\textbf{Include} \\ \textbf{HS attacks}} 
      \\
      \hline
      Edman \textit{et al.} \cite{EDMAN2009} & 2009 & Survey existing anonymous communication systems & \ik{6} & \xmark & \xmark 
      \\
      \hline
      Salo \cite{SALO2010} & 2010 & Survey Tor attacks & \ik{10} & \xmark & \xmark 
      \\
      \hline 
      Nepal \textit{et al.} \cite{NEPAL2015} & 2015 & Survey de-anonymisation attacks on hidden services & 3 & \xmark & \cmark 
      \\
      \hline 
      Erdin \textit{et al.} \cite{ERDIN2015} & 2015 & Survey de-anonymisation attacks on users & 19 & \cmark & \xmark 
      \\
      \hline 
      Yang \textit{et al.} \cite{YANG2015} & 2015 & Classification of de-anonymisation techniques & 6 & \xmark & \xmark 
      \\
      \hline 
      AlSabah \textit{et al.} \cite{ALSABAH2016} & 2016 & Survey the research on performance and security of Tor & 22 & \cmark & \cmark 
      \\      
      \hline 
      Evers \textit{et al.} \cite{EVERS2016} & 2016 & Survey Tor attacks & \ik{38} & \cmark & \cmark 
      \\
      \hline 
      Saleh \textit{et al.} \cite{SALEH2018} & 2018 & Survey all aspects of Tor research & \ik{23} & \cmark & \cmark 
      \\
      \hline 
      Aminuddin \textit{et al.} \cite{AMINUDDIN2018} & 2018 & Survey existing approaches for classifying Tor traffic & N/A & \xmark & \xmark 
      \\
      \hline 
      Kohls \textit{et al.} \cite{KOHLS2018} & 2018 & An evaluation framework for confirmation attacks & \ik{10} & \xmark & \xmark 
      \\
      \hline 
      Cambiaso \textit{et al.} \cite{CAMBIASO2019} & 2019 & Survey Tor attacks & \ik{13} & \xmark & \cmark 
      \\
      \hline 
      Basyoni \textit{et al.} \cite{BASYONI2020} & 2020 & Survey traffic analysis attacks on Tor & \ik{8} & \xmark & \cmark 
      \\
      \hline 
      Our paper & 2020 & \ik{Survey de-anonymisation attacks on Tor} & \ik{52} & \cmark & \cmark 
      \\
      \hline
    \end{tabular}
    \begin{tablenotes}
     \item \ik{WF - Website Fingerprinting, HS - Hidden Service, N/A - Not Applicable}
     \item[*] \ik{This number denotes the number of de-anonymisation attacks on Tor referenced in the paper.}
   \end{tablenotes}
   \end{threeparttable}
\end{table*}

A survey on a vast area of overall Tor research (performance, architectural improvements, attacks, and experimentation) was published in 2018 \cite{SALEH2018}. In their paper, Saleh \textit{et al.} \cite{SALEH2018} divide all Tor research into three main categories: de-anonymisation, path selection and performance analysis, and architectural improvements. The de-anonymisation category is discussed under six sub-categories: 
1. HS identification, 
2. Tor traffic identification,  
3. Attacks on Tor, 
4. Tor traffic analysis attacks, 
5. Tor improvements, and
6. Providing anonymity without Tor. 
Although Saleh \textit{et al.} describe around 23 de-anonymisation attacks under their categorisation, they have missed several important attacks, including Raptor \cite{SUN2015}, website fingerprinting attacks \cite{HERRMANN2009, HAYES2016, PANCHENKO2016} as well as most of the recent attacks. However, their paper compares Tor with other anonymity services and surveys the literature on all Tor research, focusing on experimentation, simulations, and analysis. Therefore, Saleh \textit{et al.'s} paper \cite{SALEH2018} enables readers to gain a broader knowledge of Tor research that has been conducted over the years.

Evers \textit{et al.} \cite{EVERS2016}\footnote{The work is only found on Github as it appears to be an internal university report. We could not find any published work based on this report. Also, note that this report has not been cited in any other publication previously.} report on Tor attacks known before 2016. Their report contains a corpus of references of Tor attacks, including 38 de-anonymisation attacks under their taxonomy. These attacks have been sorted into seven categories: correlation attacks, congestion attacks, timing attacks, fingerprinting attacks, DOS attacks, supportive attacks, and revealing HS attacks. This categorisation is inspired by the classification of de-anonymising techniques presented by Yang \textit{et al.} \cite{YANG2015} in 2015. 
Yang \textit{et al.} divide de-anonymising attacks into four categories based on the following two dimensions:
1. \textit{passive} and \textit{active} attacks based on the ability to manipulate traffic, and
2. \textit{single-end} and \textit{end-to-end} attacks based on the capability of the attacker to monitor or control the traffic or devices either at the sending end, receiving end, or both.
Evers \textit{et al.} try to associate this classification with their taxonomy, e.g. by classifying correlation attacks as end-to-end passive attacks, congestion and timing attacks as end-to-end active attacks, and fingerprinting attacks as single-end passive attacks. \ik{Our paper contains more de-anonymisation attacks on Tor when compared with \cite{EVERS2016}, including 17 attacks since 2016, which makes our paper the most comprehensive paper on de-anonymisation attacks to date. Moreover, we present Tor's improvements over the years and discuss how they strengthened Tor's security against de-anonymisation.}

Cambiaso \textit{et al.} \cite{CAMBIASO2019} provide a recent review of Tor attacks under a taxonomy based on the target of the attack. In this situation, the client, the server, and the network were considered targets. Although Cambiaso \textit{et al.'s} work was published in 2019, they only referenced the survey by Nepal \textit{et al.} \cite{NEPAL2015} as existing survey literature. In Cambiaso \textit{et al.'s} paper, although attacks on the Tor clients include de-anonymising the Tor user, the authors only reference less than ten such attacks. In contrast, our paper presents more than thirty such attacks. In \cite{CAMBIASO2019}, attacks on the server focus on de-anonymisation or weakening the HS, while attacks to the network consider DOS attacks and bridge discovery attacks. Moreover, Cambiaso \textit{et al.} mention some attacks under a general category in which multiple targets are considered. However, this work does not discuss sufficient details on website fingerprinting attacks - a widely researched attack in recent times.

In a recent work published in 2020, Basyoni \textit{et al.} \cite{BASYONI2020} present details on several Tor attacks from the perspective of the attack's adopted threat model. The authors of \cite{BASYONI2020} categorise their attack corpus into three threat models; a global adversary, capturing entry flows and compromising Tor nodes. They compare Tor's original threat model with the above threat models. Additionally, the practicality of these threat models is discussed in their paper. Basyoni \textit{et al.'s} paper is the only paper that has referenced a substantial amount of attacks since 2016. However, they do not reference any other survey work related to Tor attacks and only describe 8 de-anonymisation attacks in detail. Also, very little information is provided in their paper on website fingerprinting attacks and attacks on hidden services when compared with our paper.


\begin{figure*}[h]
\centering
\includegraphics[width=\textwidth]{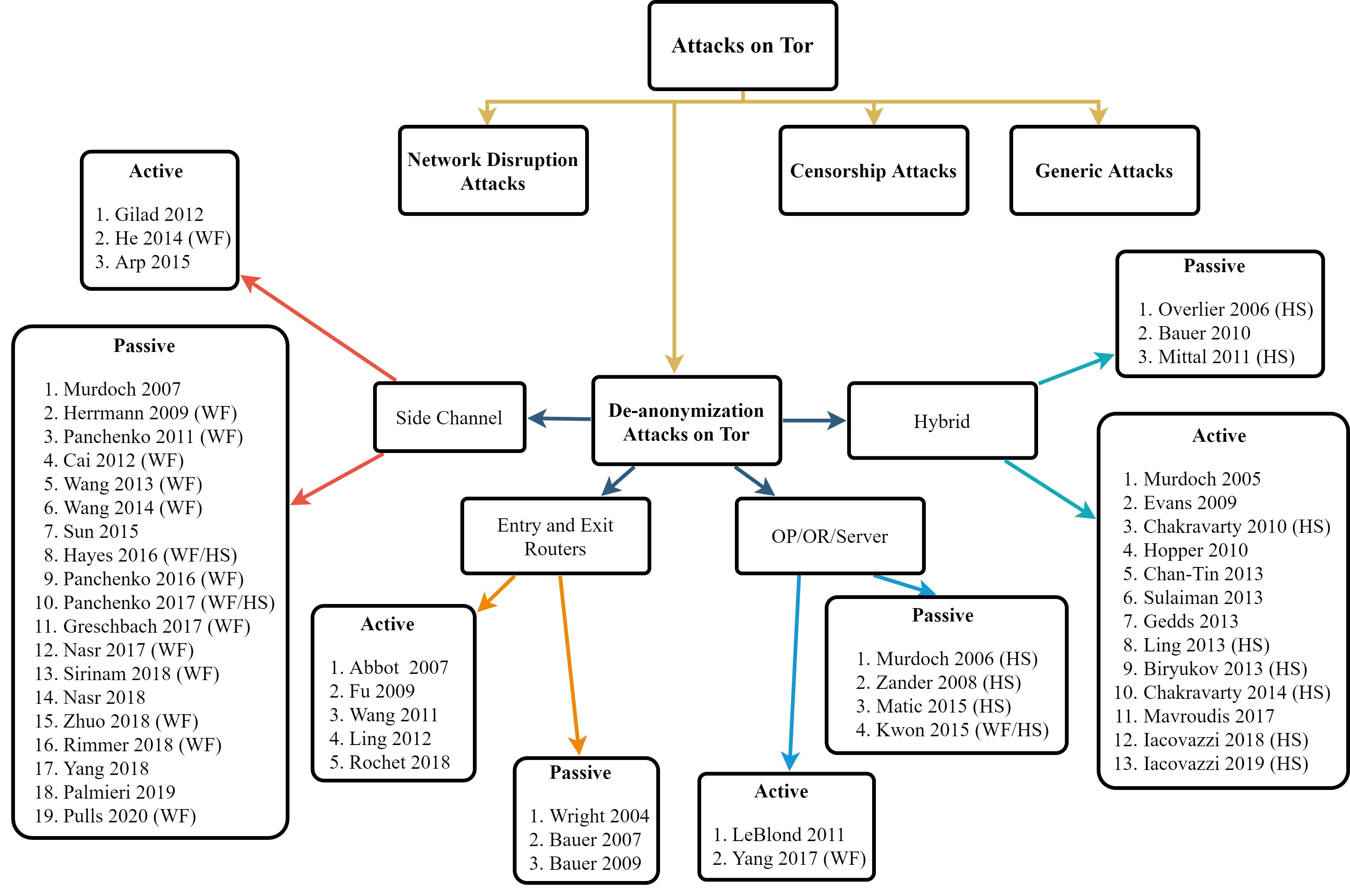}
\caption{Taxonomy for Tor Attacks}
\label{fig:Taxonomy}
\end{figure*}

Several other works have been associated with Tor attacks but focus on a different aspect of Tor research. For example, Aminuddin \textit {et al.} \cite{AMINUDDIN2018} investigate the existing literature on Tor traffic classification. Their work focuses on how machine learning techniques have been applied to such classifications and compares those techniques. The authors in \cite{AMINUDDIN2018} present a traffic classification taxonomy based on input, method, and output. The input data is categorised into circuit, flow, and packet features, while the method is categorised into supervised, semi-supervised, and unsupervised categories. The output is divided into five categories: traffic cluster, application type, application protocol, application software, and fine-grained. The last category is the one that contains the most detailed information on the classified traffic. Almost 14 previous works have been referenced in this survey. The authors in \cite{AMINUDDIN2018} claim that no classification algorithm can be presented as superior as the algorithm's efficiency and capabilities depend on the classification objective, implementation strategy, and the training dataset. Aminuddin \textit {et al.} suggest that the five factors; accuracy, training time, computational resources, number of features, and number of parameters must be considered when deciding the right algorithm for any situation.

Kohls \textit{et al.} \cite{KOHLS2018} present an analysis framework called \textit{DigesTor} for the evaluation of traffic analysis attacks. The main attack scenario considered by DigesTor is a passive attack, executed by correlating traffic features at the entry and exit of the Tor circuit. Kohls \textit{et al.} address the difficulties associated with comparing different types of traffic analysis attacks due to the diversity of the methods used.
DigesTor has two main features: 
1. a traffic analysis framework that considers five comparison metrics (namely attack type, adversary model, evaluation setup, consideration of background noise, and consideration of different application types) and estimates the similarity between the network observations, and,
2. a virtual private Tor network that is used to generate traffic for representative scenarios. 
The authors of \cite{KOHLS2018} claim that they have provided the first performance comparison of existing attacks based on their de-anonymisation capabilities.

Table \ref{table:summary} summarises the main features of the prior work discussed and highlights the significance of our paper. Although some of the previous work has included details of de-anonymisation attacks on Tor, the main focus of these papers is not to survey the attack literature, but to study a different or broader aspect of Tor, e.g. anonymity networks \cite{EDMAN2009}, performance improvements for Tor \cite{ALSABAH2016}, or broader Tor research \cite{SALEH2018}. Several other works have referenced only a small number of Tor attacks, even when their main focus is to present a survey on Tor attacks \cite{SALO2010,NEPAL2015,CAMBIASO2019,BASYONI2020}. Moreover, we have observed that most surveys do not provide much information on website fingerprinting attacks. Erdin \textit{et al.'s} \cite{ERDIN2015} work does not address attacks on HS, while Nepal \textit{et al.'s} \cite{NEPAL2015} work only focuses on the de-anonymisation of HS. None of the papers except for Saleh \textit{et al.} \cite{SALEH2018} present information on other survey work, and only Basyoni \textit{et al.} \cite{BASYONI2020} have referenced a significant number of attacks published since 2016. In this paper, we try to overcome all of these shortcomings and provide a comprehensive survey focused on de-anonymisation attacks.


\section{TAXONOMY OF TOR ATTACKS}

In this section, we present details of our proposed multi-level taxonomy of Tor attacks (see Figure \ref{fig:Taxonomy}) and explain the discriminating factor employed at each level of the taxonomy. At the top level, we use the primary objective or motive of the adversary as the differentiating factor to divide all Tor attacks into four main categories; De-anonymisation attacks, Network disruption attacks, Censorship attacks, and Generic attacks. At the middle level, we further divide the attacks based on the target component of the Tor network. At the lowest level, the differentiating factor is the methodology used by the adversary (active or passive).

\textbf{\textit{De-anonymisation attacks}}: \ik{As Tor is an anonymity network, this is the most popular type of attack against Tor. Two main scenarios are related to de-anonymisation attacks on Tor: 1. Linking (associating) client IP addresses with the IP addresses of the websites the client visits through the Tor network. For example, an entity such as an LEA might want to investigate suspicious individuals and find out what websites they are visiting over Tor. Also, LEAs might be monitoring a specific web service to identify its users. 2. Revealing the actual IP address of a HS, which is protected by Tor’s anonymity. We provide more information on de-anonymisation attacks in Section V.} 

\textbf{\textit{Network Disruption attacks}}: The main intention of these attacks is to disrupt the network, \na{usually through DOS, which makes a network unavailable for users. The attack can be launched against a single OR (a bridge or exit node, for example) or a subset of ORs or DSs.}
The CellFlood attack by Barbera \textit{et al.} \cite{BARBERA2013}, the packet spinning attack by Pappas \textit{et al.} \cite{PAPPAS2008}, and the most recent bandwidth-based DOS attacks by Jansen \textit{et al.} \cite{JANSEN2019} are some examples for these kinds of attacks. In this paper, we do not cover these attacks in detail.

\textbf{\textit{Censorship attacks}}: Tor is popularly used as a censorship circumventing tool. It allows users in oppressive and totalitarian governments to bypass censorship measures and access restricted content. Tor introduced bridges that are unadvertised relays to facilitate this. Therefore, such governments are motivated to find the means to prevent access to the Tor network. 
Attempts by various parties to block access to the Tor network are therefore considered to be Censorship attacks. China's attempts to block Tor \cite{WINTER2012}, \cite{ENSAFI2015} and blocking systems such as Nymble \cite{JOHNSON2007} are some examples of such attempts. Deep packet inspection to block Tor traffic \cite{SAPUTRA2016} and Tor bridge discovery attacks \cite{LING2012BRIDGEDISCOVERY} can also be categorised under censorship attacks.

\textbf{\textit{Generic attacks}}: \na{This is a catch-all category that encompasses many kinds of attack that are not classified elsewhere. It includes attacks such as fingerprinting attacks that identify Tor traffic \cite{LASHKARI2017} or bridges \cite{ALSABAH2016}, Sybil attacks that control a disproportionate number of nodes \cite{WINTER2016}, and Denial of Service attacks such as the Sniper attack described in \cite{JANSEN2014}. We note that most of the attacks under the generic category are a precursor to other attacks. For example, some de-anonymisation attacks require the attacker to control the entry node of the circuit. Therefore, an attack that manipulates the client to select compromised guard nodes, as discussed by Li \textit{et al.} \cite{LI2015} can be advantageous in such circumstances. Similarly, Tor traffic fingerprinting attacks are usually executed as a precursor to a censorship attack.}   
Note that we do not cover censorship attacks and generic attacks in detail in this paper.

When surveying past research efforts, we note that different terminologies are used to classify Tor attacks. We will now explain some of these terminologies and how they fit into our taxonomy.

\textit{Traffic confirmation attacks}: These are attacks in which an adversary is able to monitor both ends of a network connection (either by compromising the entry and exit nodes or by monitoring the links to and from the Tor network), to link the user and the destination. In these attacks, the adversary tries to confirm the actions of a targeted user rather than trying to uncover a random user's online activity. The confirmation attack conducted by Rochet \textit{et al.} \cite{ROCHET2018} is one of many such examples. Traffic confirmation attacks are categorised under de-anonymisation attacks in our taxonomy. 

\textit{Correlation attacks}: These attacks also come under de-anonymisation attacks. Almost all confirmation attacks require a correlation mechanism to link traffic observed at different parts of the Tor network. Different correlation techniques, such as the Spearman's rank correlation coefficient used in Raptor attacks \cite{SUN2015}, Mutual Information \cite{ZHU2004}, and Cross-correlation \cite{LEVINE2004}, can be used to identify a user's online activity from monitored traffic features.

\textit{Timing attacks}: These can be categorised as a sub-category of correlation attacks where timing characteristics in network traffic are correlated to find the link between Tor users and their online activity. The most intuitive feature used in these attacks is the inter-packet arrival time \cite{LEVINE2004}. Packet rate, used by Gilad \textit{et al.} \cite{GILAD2012}, and latency are some other features used in the timing attacks.

\textit{Watermarking attacks}: These attacks are also a form of correlation attacks where the attacker can actively manipulate the network traffic by injecting, modifying, or deleting traffic. In these attacks, a recognisable pattern is introduced into the traffic stream at one point, expecting it to be observed at another \cite{IACOVAZZI2017}.






\section{DE-ANONYMISATION ATTACKS}
\ik{Before we delve into the details of the existing research covering de-anonymisation attacks, it is pertinent to explain the selection criteria we followed to select papers to include in this review. When collecting papers for the survey, we first searched for papers on Tor attacks and related survey work\footnote{\ik{We used combinations of keywords such as Tor, survey, anonymity, attack, de-anonymisation, correlation, taxonomy, website fingerprinting, traffic analysis, timing attacks, confirmation attacks, and watermarking attacks. Next, we compiled a list of papers with attacks on Tor by going through those papers and their references. We repeated this process for every new paper we found until we could not find any new relevant papers.}}. The next step was to shortlist the most important papers to include in our survey. To do this, we went through the papers again to check their relevance to our survey and filtered papers that only focused on de-anonymisation attacks on Tor. Then, we selected the papers published on high-quality venues or papers with at least 20 citations. Finally, we checked the attack schemes of the remaining papers and selected the ones we thought would be significant.}

As previously mentioned, our primary focus is on attacks that try to de-anonymise the user, the HS, or both. De-anonymising the user is usually conducted with one of two objectives: finding out who is visiting a particular website or finding out what websites are being visited by a targeted user. Research on de-anonymisation attacks contributes to a larger portion of work carried out under Tor research. Therefore, we aim to provide an explicit categorisation and an extensive analysis of such type of attacks.

Our proposed taxonomy is multi-level. For de-anonymisation attacks, at the top level, we consider four sub-categories based on the attacker's capabilities to compromise and control network components. The network components considered are the onion proxy (Tor client), onion routers (entry, middle, exit, introduction point, rendezvous point (RP)), HS and an external web server. Any other resources that exist outside these components are considered to be side channels. Furthermore, attacks that use a mix of components are classified as hybrid. We further explain the classification below. 

\textbf{\textit{Entry and Exit routers}}: These attacks require the attacker to control both the entry and exit nodes of the circuit.

\textbf{\textit{Onion Proxy(OP)/Onion Router(OR)/Server}}: These attacks are launched by an attacker that controls a single Tor node (either a Tor client, onion router or a server). This category might seem a bit broader than the others. Still, the fact that Tor's threat model itself makes it very hard to carry out a de-anonymisation attack with a single component has been taken into consideration.

\textbf{\textit{Side channel}}: This considers attacks carried out using other means, for example, by monitoring and manipulating the links between circuit components, e.g. the link between the user and the entry node.

\textbf{\textit{Hybrid}}: This category considers a combination of components used in the above categories.

\na{At the lower level of our taxonomy (see Figure \ref{fig:Taxonomy}), we have further divided the four categories at the top level (of de-anonymisation attacks) into  \textit{active} and \textit{passive} attacks. Categorisation at this level is thus based on the methodology used by the attacker.} In passive attacks, the adversary does not modify, insert, or delete traffic on the network but can only observe and collect network traffic passively to be used in the attacks. In active attacks, the adversary manipulates network traffic in various ways to identify traffic patterns. Figure \ref{fig:Taxonomy} shows a summary of all the attacks we have categorised in this paper. Now we present details on these attacks and explain how they fit into our taxonomy.

\subsection{Entry and Exit Onion Routers}

\begin{figure}[h]
\centering
\includegraphics[width=8cm]{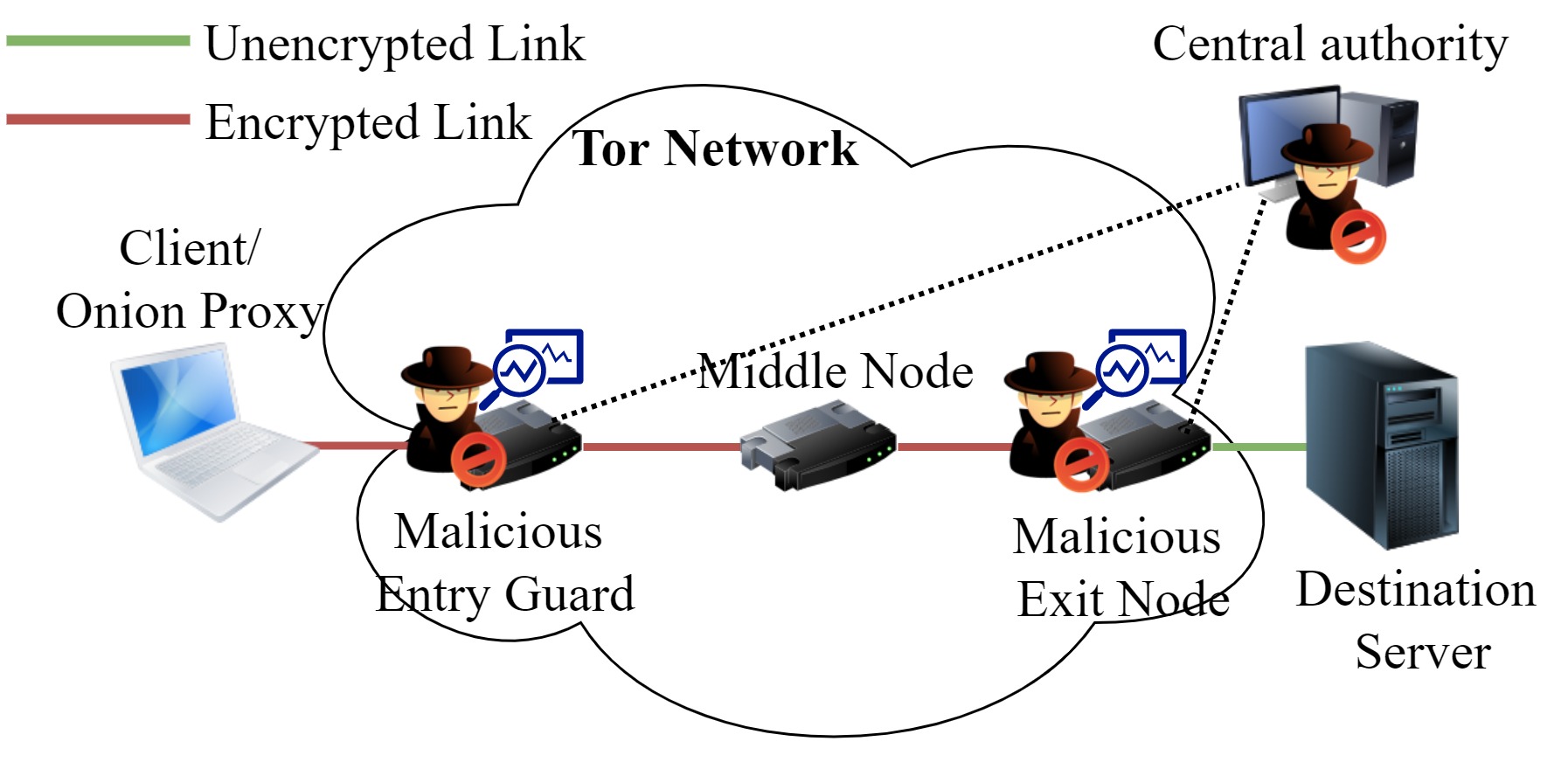}
\caption{Attack Scenario with compromised Entry and Exit nodes}
\label{fig:Attack scenario entry and exit}
\end{figure}

This category of attacks requires an adversary to access both entry and exit ORs of a Tor circuit, which can be achieved by either compromising existing Tor nodes or introducing new attacker-controlled nodes into the Tor network. When introducing new nodes, certain steps can be taken to increase the chances of a Tor node being selected as an entry or an exit node. Tor nodes can specify that they must only be used as exit nodes and configure exit policies to allow selected protocols, which improves the possibility of a particular Tor node to be chosen as an exit node. Furthermore, a node can falsely advertise high bandwidths and high uptime to be selected as an entry guard. In the early stages of the Tor network, which featured a small number of active nodes, there was a high probability of success for the attacker deployed nodes to be selected as part of circuits. Figure \ref{fig:Attack scenario entry and exit} shows the general attack scenario for this category \ik{where a malicious entry guard and a malicious exit node are nodes that are either compromised or run by the attacker. Conceptually, both of these nodes are connected to another attacker-controlled device called the central authority. However, the central authority represents a component with access to the data from both the entry and exit nodes. It can either be a completely different device or one of the nodes itself (e.g., if the entry node sends all of its data to the exit node, we can consider the exit node as the central authority). As the central authority has data from both the entry and exit nodes, it can then process the data and correlate the traffic flows to de-anonymise the client/user.}

\subsubsection{\textbf{Passive Attacks}}

Bauer \textit{et al.} \cite{BAUER2007} in 2007 described one of the earliest attacks to de-anonymise Tor circuits. Their attack is carried out in two phases. In the 1st phase, the attacker needs to control a large number of Tor relays. Either introducing new malicious nodes into the network or hijacking existing nodes in the network can achieve that. In the earlier versions of the Tor network, the router could advertise an incorrect bandwidth and uptimes to the \ik{Directory Server (DS)}. These values were not verified by the DS or the OP when selecting that particular router for a circuit. Therefore, resources required for this attack could be reduced by advertising false bandwidths for low bandwidth connections, thus increasing the chances of the adversary-controlled routers being selected as the entry or exit nodes of a circuit. Moreover, the malicious routers could have fewer restrictions on their exit policies to further increase the chances of being selected as exit nodes. If a circuit selects only one compromised relay, that relay can stop the traffic flow and force the circuit to rebuild with a different set of relays. This path disruption can be repeated until a target circuit selects two compromised routers as its entry and exit nodes. The next phase of the attack requires traffic correlation. In this phase, each malicious router in the network has to log information for each cell received, including its position in the circuit, local timestamp, previous connection's IP address and port, and next hop's IP address and port. A centralised authority that receives the above details from all malicious routers can execute a correlating algorithm to associate the sender with the receiver. 

In 2009, Bauer \textit{et al.} \cite{BAUER2009} presented their investigations on the impact of application-level protocols for the path compromising phase in \cite{BAUER2007}, which we discussed previously. In this follow-up paper \cite{BAUER2009}, it is assumed that the adversary can configure the routers with an exit policy to attract a specific application type. As an external web service can only view the IP address of the exit router, it is usually the exit router operator who is contacted when illegal activities are carried out using a Tor connection. To mitigate this abusive use of exit nodes, node operators can define \textit{exit policies}, allowing only selected services to be used and imposing several other restrictions. Due to the ability to restrict specific ports using these policies, the exit bandwidth is not uniformly distributed among different application types making some application types more vulnerable to path compromise. The results of \cite{BAUER2009} show that an adversary with control of 6 out of 1444 total routers can compromise 7.0\% of all circuits transporting Hypertext Transfer Protocol (HTTP) traffic. Meanwhile, this number is between 18.5-21.8\% for circuits that are transporting \ik{Simple Mail Transfer Protocol} (SMTP) and \ik{peer-to-peer} (P2P) file-sharing traffic.

Wright \textit{et al.} \cite{WRIGHT2004} introduced the \textit{Predecessor Attack}, a de-anonymisation attack that is applicable to many anonymity networks. In the generic predecessor attack, the attacker controls multiple nodes in the anonymity network and attempts to determine circuits consisting only of these nodes. Various techniques, such as timing analysis, are used to achieve this objective. If the complete circuit consists only of attacker-controlled nodes, the attacker can identify the user (the sender's IP address). However, it is important to note this attack is based on multiple assumptions. 1. Nodes in a path are chosen uniformly at random. 2. Repeated connections are established between the user and destination until the connection is de-anonymised. 3. Only one user maintains a session with a given destination. 4. The last node can associate a session with the target destination. When applying the predecessor attack specifically to the Tor network, the attacker only needs to control two nodes - the entry and exit nodes.

\subsubsection{\textbf{Active Attacks}}

In 2007, Abbot \textit{et al.} \cite{ABBOTT2007} published a paper describing a timing analysis attack. In their attack, a malicious exit router modified the HTTP traffic to the client by inserting an invisible \textit{iframe} that contained a JavaScript code. The Tor user's browser executed this JavaScript code, which sent regular distinctive signals to a malicious web server. As the Tor client selects a new circuit at regular intervals to increase its anonymity, if one of the malicious entry guards is selected at a certain time, the attacker can use timing analysis to de-anonymise the user. The authors of \cite{ABBOTT2007} show that even when JavaScript is disabled, this attack can be carried out using the HTML meta refresh tag, although this is more noticeable to the user. In addition, this attack can easily be executed if the user uses the \textit{Tor Button}\footnote{Before introducing the Tor Browser Bundle, Tor users had to use Firefox to access the Tor network. TorButton is an add-on for Firefox to switch the browsers' Tor usage.} to toggle the Tor proxy while keeping the browser tab open. 
This attack is more simplified when there is less traffic in a Tor connection. Abbot \textit{et al.} suggest two features that can be exploited to achieve that condition: 1. using unpopular ports and 2. maintaining the TCP stream on the circuit for more than ten minutes. This attack fails if a malicious node is not selected as an entry guard. In such situations, the attacker can use the same technique to identify entry guards and execute DOS attacks on them, forcing the client to choose a different set of entry guards. This scenario provides an opportunity for the malicious nodes to be selected as entry guards, increasing the effectiveness of this attack.

Wang \textit{et al.} \cite{WANG2011} present another active attack that utilises entry and exit nodes. For this attack, the adversary needs to control an exit router and monitor the traffic pattern at the entry guard. When the malicious exit node detects a web request of interest, it inserts a forged web page (forged web page injection attack) or alters a web page received from the server (target web page modification attack). This malicious web page causes the victim's browser to send detectable traffic patterns that the adversarial entry guard can identify to confirm the user's identity. However, this attack can be executed with only a malicious exit node if the adversary can monitor the link between OP and the Tor network. 
Wang \textit{et al.} claim that the attack is highly efficient and can identify clients using a single web request while supporting normal web browsing. This scenario provides an additional advantage to the attacker to remain undetected. Furthermore, this attack can be executed even when active content systems (e.g. JavaScript) are disabled.

A new class of active attacks, \textit{protocol level attacks}, are introduced by Fu \textit{et al.} \cite{FU2009}. These can be executed by manipulating a single cell in the circuit. These attacks need the attacker to control both the entry guard and the exit router of the circuit and need to have the ability to modify, duplicate, insert, or delete cells at the entry node. It logs the source IP address and port, circuit ID, and the time the cell was manipulated. The cell can be manipulated by duplicating and forwarding it at a suitable time, modifying a few bits of the cell, inserting a new cell into the flow, or deleting the cell. Tor uses \ik{counter mode Advanced Encryption Standard} (AES-CTR) for encryption, and when the cells are decrypted at the exit OR, the above changes in cells disrupt the counters. These disruptions cause cell recognition errors that can be observed by an attacker monitoring the exit node. The attacker records the time of these errors along with the destination IP, port, and circuit ID. Therefore, if the attacker controls both the entry and exit nodes of a circuit, he/she can use this information to correlate and link the source and the destination.

In 2012, Ling \textit{et al.} \cite{LING2012CELLCOUNTING} proposed a type of attack requiring the attacker to control a few of the Tor network's entry guards and exit nodes. This type of attack is motivated by the observation that even though Tor uses equal-sized cells at the application layer, the network's IP packets' size generally varies. The attacker selects an appropriate time and embeds a signal into the incoming traffic from the server. This task is undertaken at the exit node. An entry guard recognises this signal, a sequence of binary bits (three cells for binary ``1" and one cell for binary ``0"). However, it is possible that due to network delays and congestion, this signal may be distorted at the middle node or the links connected to it. The adversary entry node records information relevant to received cells along with the client's IP address, port, and circuit ID. Following this, the attacker decodes the embedded signal. If a match is found, the attacker is able to link the user with the destination.

The Tor protocol has a packet dropping behaviour that is common in network protocols. Rochet \textit{et al.} \cite{ROCHET2018} exploit this behaviour to launch a de-anonymisation attack, referred to as a \textit{dropmark attack} against Tor clients. In most cases, when a Tor edge node receives an unwanted cell such as a \textit{relay drop} cell or an unknown subtype of relay data cell, those cells are dropped at the edge of the circuit without tearing the circuit down. In a de-anonymisation attack, the attacker requires control of both a guard node and an exit node. Also, the circuit must have an idle time-interval. In Tor, the authors of \cite{ROCHET2018} have identified such a gap in the cell's transmission from the exit node to the client, between the \textit{connected} cell after a \ik{Domain Name System} (DNS) query, and the response of the client's GET request. The attacker sends three relay drop cells from the exit nodes during this period that are identified by the malicious guard node. 


\begin{table*}[t]
  \footnotesize
  \centering
  \caption{\ik{Summary of attacks in the first two categories}}
  \label{table:Attack summary_1}
  \begin{threeparttable}
    \begin{tabular}{ |l|l|l|l|l|l|l| } 
      \hline
      \textbf{Publication} & \textbf{Year} & \textbf{Component/s used} & \textbf{Active/Passive} & \textbf{Target} & \textbf{No of nodes in Tor \tnote{*}}
      \\
      \hline
      Wright \textit{et al.} \cite{WRIGHT2004} & 2004 & Entry and Exit & Passive & Standard Tor user & 32
      \\
      \hline
      \ik{Murodch \cite{MURDOCH2006}} & 2006 & Onion proxy & Passive & Hidden service & N/A
      \\
      \hline
      Bauer \textit{et al.} \cite{BAUER2007} & 2007 & Entry and Exit & Passive & Standard Tor user & 1344
      \\
      \hline
      Abbot \textit{et al.} \cite{ABBOTT2007} & 2007 & Entry and Exit & Active & Standard Tor user & 1344
      \\
      \hline
      \ik{Zander \textit{et al.} \cite{ZANDER2008}} & 2008 & Onion proxy & Passive & Hidden service & N/A
      \\
      \hline
      Bauer \textit{et al.} \cite{BAUER2009} & 2009 & Entry and Exit & Passive & Standard Tor user & 1880 (350 bridges)
      \\
      \hline
      Fu \textit{et al.} \cite{FU2009} & 2009 & Entry and Exit & Active & Standard Tor user & 1880 (350 bridges)
      \\
      \hline
      Wang \textit{et al.} \cite{WANG2011} & 2011 & Entry and Exit & Active & Standard Tor user & 3216 (781 bridges)
      \\
      \hline
      Le Blond \textit{et al.} \cite{LEBLOND2011} & 2011 & Exit node & Active & BitTorrent user over Tor & 3216 (781 bridges)
      \\
      \hline
      Ling \textit{et al.} \cite{LING2012CELLCOUNTING} & 2012 & Entry and Exit & Active & Standard Tor user & 4083 (1107 bridges)
      \\
      \hline
      \ik{Matic \textit{et al.} \cite{MATIC2015}} & 2015 & Onion proxy & Passive & Hidden service & N/A
      \\
      \hline
      Kwon \textit{et al.} \cite{KWON2015} & 2015 & HS Entry node & Passive & Hidden service & 9590 (2647 bridges)
      \\
      \hline
      \ik{Yang \textit{et al.} \cite{YANG2017}} & 2017 & Entry node & Active & Standard Tor user & 6110 (2104 bridges)
      \\
      \hline
      Rochet \textit{et al.} \cite{ROCHET2018} & 2018 & Entry and Exit & Active & Standard Tor user & 6862 (804 bridges)
      \\
      \hline
    \end{tabular}
    \begin{tablenotes}
     \item[*] This column shows the number of Tor nodes in the network (including bridges) by 31st of December for each corresponding year. The values were taken from the Tor project metrics \cite{TORMETRICS}. However, for the value corresponding to 2004, we have used the number given as of mid-May 2004 in the Tor deployment paper \cite{DINGLEDINE2004}. Here, we intend to highlight that due to the growth in the number of nodes in the network, these attacks require many compromised nodes to increase their probability of success in the live Tor network. \ik{As \cite{MURDOCH2006}, \cite{ZANDER2008}, and \cite{MATIC2015} use the onion proxy to execute their attacks, the number of nodes in the Tor network is not applicable (N/A) in their cases.}
   \end{tablenotes}
   \end{threeparttable}
\end{table*}

\subsubsection{\ik{\textbf{Discussion}}}

\ik{When we review the attacks that use both an entry and an exit node for execution, we can observe a few notable features. The first one is that this type of attack was prevalent in the early days of the Tor network. However, very few attacks assume access to both entry and exit Tor nodes in recent years. The growth of the Tor network and its number of users is one significant reason for this observed behaviour. As per the Tor project metrics \cite{TORMETRICS}, in December 2020, there were about 6800 live relays and 1550 bridges in operation, supporting about 2.25 million Tor users (excluding users connecting through Tor bridges). However, in 2012, there were only about 3000 relays and 1100 bridges supporting 0.8 million Tor users. These numbers make it difficult for a real-life adversary to execute an attack by deploying its own Tor relays, hoping that a user may select both entry and exit nodes controlled by the adversary in a Tor circuit. The probability of this is extremely low unless the attacker is a highly resourceful global entity. We can assume that early research on Tor focused more on breaking its anonymity by compromising nodes (referred to as structural attacks \cite{BACKES2016}), as Tor’s threat model itself focuses on protecting against an adversary that has access to a fraction of the network. Quantification of anonymity can be insightful in designing more advanced structural attacks, which might be effective against the current Tor network \cite{BACKES2016}. Furthermore, it would be interesting to explore the effect of integrating recent advancements in technology (e.g. deep learning) with these types of attacks to make them more effective.}

\ik{We also observed that active attacks are more diverse and effective as, even in 2018, Tor has demonstrated vulnerabilities that could be exploited to execute an active de-anonymisation attack \cite{ROCHET2018}. Tor’s fixed-size cells, encryption, network jitter, and other delays can affect the accuracy of passive correlation attacks. However, adding signals into the traffic, for example by modifying webpages, injecting cells, and dropping cells, can be more effective for the success of de-anonymisation attacks. The major drawbacks of active attacks are the necessity of additional resources and the possibility of being spotted easily by a more cautious user. An interested reader can go through early attacks \cite{BAUER2007,ABBOTT2007} and then refer to more recent attacks \cite{ROCHET2018} to evaluate their applicability to the current Tor network.}

\subsection{Onion Proxy/Onion Router/Server}

In this category, the adversary uses a single Tor network component such as the OP, server, or an OR. Suppose the OP (the Tor client) is used to execute the attack. In that case, its default functionality will usually be altered to match the requirements of the attack, e.g. sending periodic traffic patterns. If the attack requires a compromised server, this can be accomplished by hosting a server or taking control of a targeted server. An OR can be compromised in the same way as was explained in the previous section. Table \ref{table:Attack summary_1} summarises the attacks that fit into the first two categories and shows how the Tor network has scaled with time, making it challenging to execute attacks by controlling Tor nodes.

\subsubsection{\textbf{Passive Attacks}}

In 2006, an attack to de-anonymise HSs was published by Murdoch \cite{MURDOCH2006}. The attack described in Murdoch's paper does not require any node in the circuit to be controlled, and the attacker cannot observe, modify, insert, or delete any network traffic. However, the attacker needs to control the client in order to execute this attack. \ik{It is possible to execute this attack as different machines have different clock skews (even identical models). Clock skew is the ratio between actual and nominal clock frequencies.} The attack is executed by accessing the HS with varying traffic that affects the clock skew of the machine hosting the service. By requesting timestamps, these changes to the clock skew can be captured. The attacker then probes all suspecting machines for their timestamps. Finally, the attacker is able to reveal a correlation between the clock skew and the traffic pattern, thus de-anonymising the HS. In 2008, Zander \textit{et al.} \cite{ZANDER2008} improved upon Murdoch's \cite{MURDOCH2006} attack by focusing on the quantisation noise that limited the effectiveness of the attack. Clock skew has two main sources of noise, namely network jitter and timestamp quantisation error. Zander \textit{et al.} show that this quantisation error can be significantly minimised by synchronised sampling, reducing the impact on clock frequency.

\textit{CARONTE} is a tool that can de-anonymise HS by using location leaks in their content and configuration. This tool was developed by Matic \textit{et al.} \cite{MATIC2015}. Their approach consists of 3 steps. 1. \textit{Exploration}, which takes a set of onion URLs as input and extends each of them to include a root page, all resources, and one random resource that is added to trigger a \textit{``not found"} error page. Following this, all onion URLs in the extended set are visited through Tor, using HTTP and HTTPS to collect the HS's content and certificate chain. 2. \textit{Candidate selection}, in which a list of candidate pairs is generated using the collected information. A candidate pair consists of an onion address and an internet endpoint (either an IP address or a DNS domain). These are generated by examining endpoints, unique strings, and HTTP certificates of collected onion pages. 3. \textit{Validation}, in which CARONTE verifies whether a candidate endpoint hosts the HS. This is done by visiting the endpoints separately to collect their content and certificates and finally comparing them with those of the onion address.
This approach does not rely on a weakness in the Tor network but exploits sensitive information embedded in a HS's content and configuration. This attack only uses OP for its execution.

Kwon \textit{et al.} \cite{KWON2015} propose a \textit{circuit fingerprinting} attack to identify HSs. For this attack, the attacker needs to extract circuit-level information, such as the number of incoming and outgoing cells, sequence of packets, lifetime, and timing information. Although under certain conditions, a network administrator or an \ik{Internet Service Provider (ISP)} can obtain this information, the most realistic and effective way to execute the attack described in \cite{KWON2015} is to control an entry guard. Firstly, Kwon \textit{et al.} discuss how certain distinctive circuit features such as incoming and outgoing cells, duration of activity, and circuit construction sequences can be used to classify a given circuit into five different categories: 1. HS - Introduction point, 2. Client - RP, 3. Client - Introduction point, 4. HS - RP, and 5. General Tor circuits. Subsequently, they discuss how website fingerprinting can be used in conjunction with the circuit classification to de-anonymise a HS. According to their paper, to obtain training data from the HS side, the attacker first downloads the content from different HSs and then starts up a HS with this downloaded data in a sub-directory. The attacker's objective is to link a given network trace with a HS by using website fingerprinting techniques. Then, by using the circuit classification technique described in \cite{KWON2015}, the attacker can determine whether the trace belongs to the client-side or server-side. If it belongs to the server-side, the IP address of the HS can be identified.

\subsubsection{\textbf{Active Attacks}}

Le Blond \textit{et al.} \cite{LEBLOND2011} describe two attacks for de-anonymising Tor users by exploiting insecure applications. One attack requires the adversary to control an exit node and a publicly connectable BitTorrent peer. In this attack, the BitTorrent tracker's response is hijacked by the malicious exit node, and the IP address and port of the malicious peer are inserted into it. If the user connects to the peer directly, without using Tor, the attacker can trace the user easily. The authors of \cite{LEBLOND2011} claim that a majority of BitTorrent users use Tor only to connect to the centralised tracker. By comparing the publicly available IP addresses of the exit nodes with the IP addresses connected to the malicious peer, it is possible to verify their claim.
Distributed Hash Table (DHT) tracking, carried over the User Datagram Protocol (UDP), is exploited in the second attack. As Tor only supports \ik{Transmission Control Protocol (TCP)}, the BitTorrent client cannot connect to DHT using Tor; however, DHT keeps track of the IP addresses and ports of peers downloading specific content. This type of attack is carried out when the exit node identifies a target user connecting to the BitTorrent tracker via Tor. The content identifier and the port number for a specific download are included in the BitTorrent subscription to the tracker and the handshake messages. Following this, the attacker tries to match a user with a similar port number from the list of candidate IP/ports for that specific content ID in the DHT. If the attacker finds a match, then the Tor user can be de-anonymised.
In addition to the above attacks, Le Blond \textit{et al.} present an attack, known as the \textit{Bad apple attack}, which can be used to identify the IP address of other streams once a BitTorrent stream is de-anonymised. The fact that all streams multiplexed into the same circuit originate from the same user can be used to de-anonymise the other streams in that circuit. When it comes to different circuits, the attack exploits two BitTorrent signalling patterns; 1. the peer identifier (can only be used if the peer-to-peer communication is not encrypted), and 2. the IP address and the port returned in the tracker response. Therefore, two circuits can be linked if a peer in one circuit communicates with an IP/port included in the tracker response of another circuit. The authors of \cite{LEBLOND2011} were also able to use this technique to trace HTTP streams.

\ik{Yang \textit{et al.} \cite{YANG2017} propose a new active website fingerprinting attack, following up on the work of He \textit{et al.} \cite{HE2014}. In contrast with the attack scenario in \cite{HE2014}\footnote{\ik{In He \textit{et al.'s} attack, they assume that the attacker only has access to the traffic between the Tor client and the entry node (e.g. network administrator) and therefore we have included details of \cite{HE2014} under active side-channel attacks.}}, the authors of \cite{YANG2017} assume that the attacker can control the entry node of a Tor circuit and actively manipulate the traffic going through it. In summary, an attacker attempts to delay HTTP requests by a client to retrieve web objects (e.g. images, CSS files) in order to infer information about web pages and use them in a website fingerprinting attack. Yang \textit{et al.} argue that the work in \cite{HE2014} was not able to identify HTTP requests accurately and filter control packets, while the attack in \cite{YANG2017} can address these limitations. The technique in \cite{YANG2017} consists of five steps. In the first step, Yang \textit{et al.} try to identify the first HTTP request by observing the cell transmission patterns for circuit and stream establishment. The second step is to identify the delayed position for subsequent requests for which they have designed a \textit{Cell-Delay Position Decision Algorithm} and a \textit{Delay-Scheduling Algorithm}. In the third step, they capture and record all relay cells while extracting features and creating a fingerprint in the fourth step. Finally, they use Support Vector Machine and K-Nearest Neighbour algorithms for the classification. They claim that their methods obtained a maximum accuracy of 98.64\%.}

\subsubsection{\ik{\textbf{Discussion}}}

\ik{Tor provides strong anonymity against attackers with limited resources (e.g. access to only a Tor client or router). However, it is important to note that there has been some interesting work on such attacks published in recent years \cite{KWON2015,MATIC2015, YANG2017}. There are some key details we can observe in these attacks. Most passive attacks in this category have tried to de-anonymise hidden services. As a hidden service connection does not exit the Tor network, it is imperative to figure out ways that do not involve traffic correlation, discussed mainly in the previous category. Another important feature of this category is that the attacks are unique to each other (except for \cite{ZANDER2008}, which is an improvement on \cite{MURDOCH2006}). Techniques used in these attacks vary from using clock skew \cite{MURDOCH2006}, exploiting vulnerabilities in BitTorrent application \cite{LEBLOND2011}, exploiting location leaks in HSs \cite{MATIC2015}, and executing passive and active website fingerprinting attacks on HSs, in \cite{KWON2015} and \cite{YANG2017} respectively. Some of the attacks utilise machine learning techniques (\cite{KWON2015,YANG2015}), which can set a precedent for future attacks that use a single component of the Tor circuit.}



\subsection{Side Channels}

Side-channel attacks use means other than compromising the main Tor components to execute the attack. The most common type of side-channel used against Tor is the traffic intercepted between the Tor client and the entry node. Network administrators or Internet Service Providers (ISPs) can monitor this traffic.

\subsubsection{\textbf{Passive Attacks}}

In 2007, Murdoch \textit{et al.} \cite{MURDOCH2007} addressed the ability of adversaries controlling Internet Exchange Points (IXPs) to execute passive correlation attacks. This type of attack assumes that 1. traffic going in and coming out of the Tor network for a targeted flow passes through an attacker-controlled IXP, 2. the packet sampling is distributed over the flow independently and identically, and 3. the attacker can distinguish Tor traffic from regular network traffic. The attacker then tries to match the target flow going into the network with the traffic flow coming out of the network, or vice versa. A Bayesian approach is used to infer the best possible match. Simulations have been carried out to evaluate the attack by varying the number of flows, sampling rate, mean network latency, and the attack method. In 2013, Johnson \textit{et al.} \cite{JOHNSON2013} discussed the realistic nature of this type of adversary in their paper.

A novel set of attacks, known as \textit{RAPTOR} attacks, which can be launched by \ik{Autonomous Systems (ASs)}, were presented by Sun \textit{et al.} \cite{SUN2015} in 2015. These attacks can either be executed individually or combined for improved performance.
1. Asymmetric traffic analysis - This attack considers the natural asymmetry of the internet paths and shows that anonymity network users can be de-anonymised by observing only a single traffic direction at both communication endpoints.
2. Exploiting natural churn - This attack is based on the fact that internet paths fluctuate over time due to changes in physical topology.
3. \ik{Border Gateway Protocol (BGP)} hijacking attack - This attack is also known as prefix hijacking attack. In this attack, a malicious AS advertises false BGP control messages in order to capture a part of the traffic to the victim. This captured traffic is then used to learn the IP address of the guard relay.
4. BGP interception attack - This is also called prefix interception attack. In this attack, the malicious AS becomes an intermediate AS on the Internet path. Here, the connection is kept alive, unlike in the hijacking attack, which enables asymmetrical traffic analysis. Sun \textit{et al.} also point out that the adversary can execute an interception attack at both the guard relay and exit relay simultaneously. They execute a real-world BGP interception attack against a live Tor relay by collaborating with AS operators.
These attacks exploit the dynamics of internet routing, such as routing symmetry and routing churn.

More recently, researchers have focused on using deep learning techniques to execute de-anonymisation attacks on the Tor network. Nasr \textit{et al.} \cite{NASR2018} demonstrate a traffic correlation attack by using deep learning. In their attack, a correlation function tailored to the Tor network is learned and used by the \textit{DeepCorr} system to cross-correlate live Tor flows. This system can correlate two ends of the connection - even if the destination has not been used in the training process - as its correlation function can link arbitrary circuit flows and flows to arbitrary destinations. DeepCorr's neural network learns generic features of noise in Tor, allowing it to correlate circuit flows that are different to those used during the learning process. Furthermore, DeepCorr's performance improves with higher observation times and larger training datasets. 

Palmieri \cite{PALMIERI2019} presents a flow correlation attack based on wavelet multi-resolution analysis. This is a passive attack in which the adversary must eavesdrop on ingress and egress traffic. \textit{Wavelets} are functions that satisfy certain mathematical properties and are used to represent data or other functions \cite{GRAPS1995}. Wavelet analysis can be used to obtain a clearer and more complete view of a signal, its generation and other less evident dynamics by decomposing the signal on multiple scales. Properties that are not evident by direct observation are thus identified and used to correlate the captured flows.

\begin{figure}[h]
\centering
\includegraphics[width=8cm]{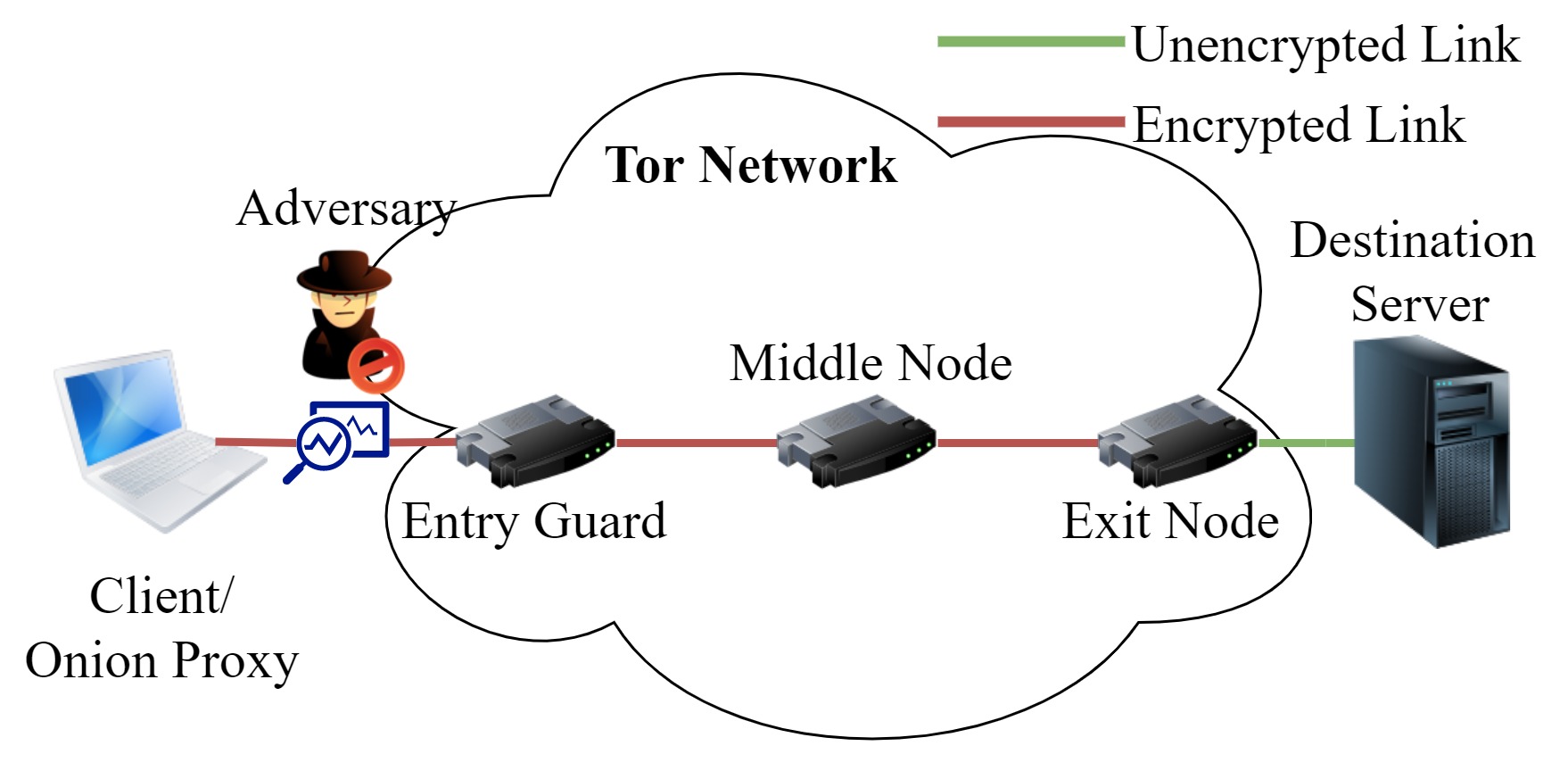}
\caption{Attack Scenario for Website Fingerprinting}
\label{fig:website fingerprinting}
\end{figure}

Concepts related to Website Fingerprinting (WF) were first explored in the late '90s. The term, \textit{Website Fingerprinting}, was coined by Hintz in 2002 \cite{HINTZ2002}. In 2009, Herrmann \textit{et al.} \cite{HERRMANN2009} presented a WF attack on the Tor network in which the adversary was able to monitor the traffic between the privacy-enhancing system and the user. This is a passive attack based on the Multinomial Naive Bayes classifier. This type of attack consists of two phases: namely, the training phase and the testing phase. In the training phase, traffic fingerprints are created for either a large number of generic sites or a small number of \textit{targeted sites}. These are then stored in a database with the corresponding URLs. In the testing phase, fingerprints are created from the traffic recorded from users and are then compared with the database records to find any matching fingerprints. Figure \ref{fig:website fingerprinting} shows the attack scenario for generic WF attacks, \ik{where the adversary is either running an entry guard or sniffing the traffic between the client and the entry guard.}

In 2011, Panchenko \textit{et al.} \cite{PANCHENKO2011} presented another WF attack based on Support Vector Machines (SVM). They define features based on volume and time, while previous WF attacks only considered packet sizes and the direction of packets (incoming or outgoing). The experiments were carried out in a closed world scenario and were later extended to an open-world scenario. In a closed world scenario, the attacker is aware of all the web pages ordinarily visited by the victim, while this is not the case in the open world. Panchenko \textit{et al.} also present preliminary results on how camouflage affects these attacks. A randomly chosen web page is simultaneously loaded with the requested web page to achieve the camouflage effect in their work.

Cai \textit{et al.} \cite{CAI2012} describe a \textit{web page} fingerprinting attack as well as a \textit{website} fingerprinting attack. Both these attacks assume that an adversary can monitor a user's internet connection. 
For the web page fingerprinting attack, network traces are converted into strings, and the \textit{Damerau-Levenshtein Distance}\footnote{Damerau-Levenshtein distance is a metric that evaluates the distance between two strings to compute their dissimilarity to each other.} (DLD) is calculated. Following this, a SVM classifier with a distance-based kernel is used for classification. This technique is then extended to a website classifier by using the Hidden Markov Models. These models help the attacker to conclude whether a sequence of web pages are from the same website or not.
Cai \textit{et al.} have evaluated their datasets against the work of Panchenko \textit{et al.} \cite{PANCHENKO2011} and Herrmann \textit{et al.} \cite{HERRMANN2009} and claim that the proposed attack mechanism in \cite{CAI2012} is far more effective. Furthermore, Cai \textit{et al.} claim that their work is the first to evaluate the security provided by application-level defences such as HTTPOS \cite{LUO2011} and random pipelining \cite{PERRY2011}, while the previous attacks only considered packet padding and other network-level defences.

In subsequent years, several works have been published on WF attacks. In 2013, Wang \textit{et al.} \cite{WANG2013} published the results of a WF attack that used a SVM classifier and two new distance-based metrics to compare packet traces. Their paper demonstrates that one metric - introduced as the \textit{combined OSAD} (Optimal String Alignment Distance) - reduces the error rate while the other metric - presented as the \textit{fast Levenshtein-like algorithm} - significantly reduces the training time. Wang \textit{et al.} follow up on the work of Cai \textit{et al.} \cite{CAI2012}, comparing each other's work. When reviewing Cai \textit{et al.'s} code and results, Wang \textit{et al.} note that the metric used by Cai \textit{et al.} in their work was actually the OSAD (which is a more restricted version of DLD) and not the DLD. Wang \textit{et al.'s} attacks are evaluated in a closed world scenario as well as an open-world scenario.

Again in 2014, Wang \textit{et al.} \cite{WANG2014} published a WF attack with a local passive adversary, applying the k-Nearest Neighbour (k-NN) classifier to a large feature set with weight adjustments. This feature set included general features such as total transmission size, the numbers of incoming and outgoing packets, as well as unique features such as packet lengths, packet ordering, the concentration of outgoing packets, and bursts. Wang \textit{et al.'s} \cite{WANG2014} paper also discusses WF defences and claims that all previous defences only work against specific attacks. Hence its authors propose a \textit{provably effective defence}, which has the capability to defeat any attack and requires fewer resources. 

Hayes \textit{et al.} introduced K-fingerprinting, a novel WF technique based upon random decision forests \cite{HAYES2016}. In their paper, the authors evaluate this type of attack against standard web pages as well as HSs. This type of attack assumes a passive attacker who can observe the client's encrypted traffic. It consists of two stages. In the first stage, the attacker captures network traffic generated from a set of web pages that he/she wishes to monitor and a large number of other unmonitored web pages and uses these traces to train a random forest for classification. Following this, the attacker captures traces from the client's browsing session. In k-fingerprinting, random forests are used to extract a fixed-length fingerprint rather than directly using the classification. Therefore, after capturing all traces, the attacker computes the fingerprint's Hamming distance from the client's traffic with the set of fingerprints collected for classification in stage one. 

In 2016, Panchenko \textit{et al.} \cite{PANCHENKO2016} published another WF attack using a passive eavesdropper that served to monitor the traffic between the client and the entry node. This fingerprinting approach was named CUMUL and required the use of an SVM classifier. Panchenko \textit{et al.'s} paper mentions three limitations of previous datasets;
1. the previous datasets contain only index pages,
2. they do not allow an evaluation of fingerprinting for complete websites, and 
3. small datasets do not allow for generalisation as the world wide web consists of billions of web pages.
Therefore, the authors in \cite{PANCHENKO2016} use novel datasets to overcome these issues. 

Following Kwon \textit{et al.'s} work \cite{KWON2015} in 2015, in 2017, Panchenko \textit{et al.} \cite{PANCHENKO2017} published their work on de-anonymising HSs using fingerprinting techniques. However, for the approach taken by Panchenko \textit{et al.}, the attacker does not need to control an entry guard but instead needs to have the ability to observe the link between the client and the guard. Their technique consists of two phases. In the first phase, they try to detect whether there is communication with the HSs. They further break down this phase into detecting unknown HS communications and detecting known HS communications. For both of these scenarios, the authors of \cite{PANCHENKO2017} apply a binary classifier. In the second phase, Panchenko \textit{et al.} try to detect the HS visited by the client. Assuming that HS communication has already been detected in phase 1, phase 2 is explored by the authors in the following two ways: 1. the adversary wants to detect a targeted set of HSs, and 2. the adversary knows all the HSs and wants to find out which one is being visited by the client. However, Panchenko \textit{et al.} claim that in general, neither this attack nor other existing attacks scale in realistic settings. The claim that WF attacks are not properly scaled to be effective in the live network had previously been discussed by Juarez \textit{et al.} \cite{JUAREZ2014} in 2014.

One of the main disadvantages of any traffic-analysis based attack is the huge storage and computational overheads required for massive traffic volumes. Nasr \textit{et al.} \cite{NASR2017} address this issue by introducing \textit{compressive traffic analysis}, a technique in which traffic analysis is conducted on compressed features and not raw features. This approach is inspired by \textit{compressed sensing} \cite{ELDAR2012}, which is an active area in signal processing. 
Nasr \textit{et al.} present two main reasons for the feasibility of compressive traffic analysis. 1. Traffic features such as packet timings and sizes are \textit{sparse} signals, in which the compressed sensing algorithms work best. 2. The \textit{restricted isometry property} of compressed sensing algorithms \cite{ELDAR2012} allows traffic features to keep their \textit{Euclidean distances} after compression, which allows traffic analysis to be conducted on compressed features. 
Based on this concept, \textit{compressive flow correlation} and \textit{compressive website fingerprinting} are introduced and compared with state-of-the-art techniques. The authors of \cite{NASR2017} used k-NN and SVM classifiers for the website fingerprinting attack, thus demonstrating that compressive website fingerprinting requires lower storage and computational time than its traditional counterparts.

An inherent issue of WF attacks in the literature is that they often neglect realistic scenarios. For example, researchers often assume that there are only discrete visits to webpages, thus ignoring hyperlink transmissions. Zhou \textit{et al.} \cite{ZHUO2018} propose a WF attack based on the Profile Hidden Markov Model (PHMM), a technique that is used for DNA sequencing analysis in bioinformatics. Their main argument is that even though there may be noise impacting the results in scenarios such as subsequent visits by a user to a webpage using hyperlinks, key elements can still be used to identify a website. The authors of \cite{ZHUO2018} equate this to the fact that while there are different genes in an organism under different environmental factors, its essential functionality genes do not change. Based on this argument, Zhou \textit{et al.} claim that their WF attack is more applicable in practice. Furthermore, their paper provides a useful taxonomy, and a comparison of WF attacks up to 2016.

\begin{table*}[t]
  \footnotesize
  \centering
  \caption{Summary of website fingerprinting attacks against Tor}
  \label{table:Attack summary_2}
  \begin{threeparttable}
    \begin{tabular}{ |l|l|l|l|l|l|l| } 
      \hline
      \textbf{Publication} & \textbf{Year} & \textbf{Classifier} & \textbf{Setting\tnote{*}} & \textbf{Unique/Novel claims}
      \\
      \hline
      Herrmann \textit{et al.} \cite{HERRMANN2009} & 2009 & Multinomial Naïve-Bayes & Closed & Apply WF to anonymous networks
      \\
      \hline
      Panchenko \textit{et al.} \cite{PANCHENKO2011} & 2011 & Support Vector Machine & Both & First successful attack for open-world scenario
      \\
      \hline
      \multirow{2}{*}{Cai \textit{et al.} \cite{CAI2012}} & \multirow{2}{*}{2012} & Support Vector Machine & Closed & Propose both \textit{web page} and \textit{website} fingerprinting attacks
      \\
      \cline{3-4}
       & & Hidden Markov Model & Open & First evaluation of application level defences
      \\
      \hline
      Wang \textit{et al.} \cite{WANG2013} & 2013 & Support Vector Machine & Both & Propose 2 new distance based metrics to compare packet traces
      \\
      \hline
      Wang \textit{et al.} \cite{WANG2014} & 2014 & k-Nearest Neighbour & Open & Design a provably effective defence against WF attacks
      \\
      \hline
      \ik{He \textit{et al.} \cite{HE2014}} & 2014 & Support Vector Machine & Closed & First active WF attack
      \\
      \hline
      Kwon \textit{et al.} \cite{KWON2015} & 2015 & k-Nearest Neighbour & Both & Apply WF for hidden services
      \\
      \hline
      Hayes \textit{et al.} \cite{HAYES2016} & 2016 & Random decision forests & Both & Possible to launch WF attacks even with lots of noise
      \\
      \hline
      Panchenko \textit{et al.} \cite{PANCHENKO2016} & 2016 & Support Vector Machine & Both & Web page fingerprinting in not practical at internet scale
      \\
      \hline
      Panchenko \textit{et al.} \cite{PANCHENKO2017} & 2017 & Binary classifier \& SVM & Both & Performance and limits of fingerprinting attacks on HSs
      \\
      \hline
      Greschbach \textit{et al.} \cite{GRESCHBACH2017} & 2017 & Modified k-NN  & Both & Associate DNS traffic for WF attacks
      \\
      \hline
      Nasr \textit{et al.} \cite{NASR2017} & 2017 & k-NN \& SVM  & Open & Introduce compressive website fingerprinting
      \\
      \hline
      \ik{Yang \textit{et al.} \cite{YANG2017}} & 2017 & k-NN \& SVM & Both & Address the limitations of \cite{HE2014} in identifying HTTP requests
      \\
      \hline
      Zhou \textit{et al.} \cite{ZHUO2018} & 2018 & Profile Hidden Markov Model & Both & Consider hyperlink transitions made by users
      \\
      \hline
      Rimmer \textit{et al.} \cite{RIMMER2018} & 2018 & SDAE, CNN \& LSTM & Both & Apply DL algorithms for WF and automatic feature extraction
      \\
      \hline
      Sirinam \textit{et al.} \cite{SIRINAM2018} & 2018 & Convolutional Neural Network & Both & Undermine WF defences considered for deployment in Tor
      \\
      \hline
      Pulls \textit{et al.} \cite{PULLS2020} & 2020 & Convolutional Neural Network & Open & Introduce the notion of a \textit{Website Oracle}
      \\
      \hline
    \end{tabular}
    \begin{tablenotes}
      \item \ik{\textit{WF} - Website Fingerprinting, \textit{SVM} - Support Vector Machine, \textit{k-NN} - k- Nearest Neighbour, \textit{DNS} - Domain Name System, \textit{HS} - Hidden Service,} \textit{SDAE} - Stacked Denoising Autoencoder, \textit{CNN} - Convolutional Neural Network, \textit{LSTM} - Long Short-Term Memory, \textit{DL} - Deep Learning
      \item [*] \ik{\textit{Setting} denotes whether the authors have conducted their experiments in a closed world scenario or an open world scenario.}
    \end{tablenotes}
   \end{threeparttable}
\end{table*}

Different WF attacks use different classifiers or feature sets. Most of the time, these features are manually extracted and are specific to a particular attack. A paper published by Rimmer \textit{et al.} \cite{RIMMER2018} in 2018 claims to present the first WF approach that carries out automatic feature extraction. Rimmer \textit{et al.} argue that since the classifier and the features are fixed for most of the attacks, it is easy to develop defences against them. However, this is not the case against their attack. Deep learning models such as the feedforward Stacked Denoising Autoencoder (SDAE), Convolutional Neural Networks (CNN), and recurrent Long Short-Term Memory (LSTM) were applied to their approach.

Sirinam \textit{et al.} \cite{SIRINAM2018} describe a more recent WF attack against the Tor network, titled \textit{Deep Fingerprinting}, which uses CNNs. They claim that this attack has an accuracy of 98\% without defences, more than 90\% against WTF-PADs \cite{JUAREZ2016}, and 49.7\% against Walkie-Talkies \cite{WANG2017}; two prominent types of defences against WF attacks that were seriously being considered for deployment by the Tor project \cite{TORPROJECT}. \ik{WTF-PAD is an adaptive padding technique in which padding is implemented when a channel is not being used much. This technique helps to conceal traffic bursts and other features that can be used to identify traffic flows. Meanwhile, Walkie-Talkie causes the server and the client to send non-overlapping bursts while adding dummy packets and delays to generate collisions. These collisions help to create similar features for multiple sites, thus protecting against ML-based classification techniques. The attack described in \cite{SIRINAM2018} was conducted in both closed and open-world settings.}

Greschbach \textit{et al.} \cite{GRESCHBACH2017} present a new set of correlation attacks called \textit{DefecTor attacks} that use DNS traffic for precision improvement. As DNS uses the User Datagram Protocol (UDP), which Tor does not support, Tor provides a workaround. The OP transfers the hostname and port to the exit node, and the exit node resolves the address. If the DNS resolution is successful, the exit node opens a new TCP connection to the target server. In Greschbach \textit{et al.'s} paper, a conventional WF attack is combined with the egress DNS traffic observed by a passive attacker. The attack can be carried out by observing the links or running a compromised entry node and a DNS resolver or server. Two DefecTor classifiers are proposed by extending Wang \textit{et al.'s} k-NN classifier \cite{WANG2014}.

In a very recent publication, Pulls \textit{et al.} \cite{PULLS2020} introduce the security notion of a \textit{Website Oracle \ik{(WO)}}, which can be combined with a WF attack to increase its effectiveness. A WO provides information on whether a particular monitored website was visited via Tor at a given time. In general, WO further improves the WF classification's performance. DNS resolvers used by Greschbach \textit{et al.} \cite{GRESCHBACH2017} are an example of a WO source. However, Pulls \textit{et al.} also mention other WO sources. Web server access logs, content delivery networks, exit relays, Tor DSs, Real-Time Bidding (RTB) \cite{JUNWANG2017}, and dragnet surveillance programs are some such examples. For their experiments, the authors of \cite{PULLS2020} used Sirinam \textit{et al.'s} Deep Fingerprinting attack \cite{SIRINAM2018} in conjunction with WOs. Table \ref{table:Attack summary_2} provides a summary of all the WF attacks we have discussed in this paper.

Yang \textit{et al.} \cite{YANG2018} worked on an entirely different de-anonymisation scenario to previous attacks we have described in this section. In their attack, they tried to identify the websites visited by a smartphone user via Tor. A malicious USB charging device, such as the ones in public USB charging stations, was the assumed attacker. Yang \textit{et al.} used the official Tor apps on Android - Orbot and Orfox. 1. Orbot implements a local proxy to provide Tor access to mobile phones, and 2. Orfox is a Firefox-based browser for smartphones. For their attack, the authors of \cite{YANG2018} considered some realistic factors such as the network type (LTE or Wifi) and the battery level. They extracted time and frequency features from these traces and used them in a random forest classifier. However, as Yang \textit{et al.} only used 100 websites (50 regular and 50 onion services) in their experiments, they have claimed their work as a proof of concept.

\subsubsection{\textbf{Active Attacks}}

Most side-channel attacks available in the surveyed literature are passive. However, Gilad \textit{et al.} \cite{GILAD2012} describe an active attack where the attacker influences the rate of communication between the exit node and the server and is, therefore, able to observe the traffic between the client and the entry guard. Firstly, the attacker sends spoofed packets (with the server's address and port as a source) from the probe circuit to the exit node. Then the exit node sends a duplicate acknowledgement (ACK) to the server. TCP interprets three such duplicate ACKs as a congestion event, and the servers' congestion window shrinks, resulting in a reduction of the transmission rate. The attacker observes this at the client end, thus de-anonymising the communications. However, it should be noted that Gilad \textit{et al.'s} paper \cite{GILAD2012} was published with only preliminary results for this type of de-anonymisation attack.

\ik{In 2014, He \textit{et al.} \cite{HE2014} proposed an active WF attack. They argue that the overlapping of web objects (e.g. images, CSS files) in returned web pages affects the accuracy of passive WF attacks. When a browser parses a HTML document, it needs to send requests to retrieve different web objects. If these web objects are larger than the maximum transmission unit, they will be split and transmitted in multiple packets. This scenario will produce multiple incoming packets (packets from the server to the client) just after a single outgoing packet (packets from client to server), a scenario which is defined by He \textit{et al.} as a burst. They argue that features related to these bursts give insights into the size of the web objects on a web page and provide higher accuracy for WF attacks. However, incoming packets from multiple HTTP requests usually overlap, impacting the effectiveness of those features. If the outgoing requests can be delayed until the previously requested web object is fully downloaded, the above issue can be overcome. This phenomenon is the main idea behind the active WF attack introduced by He \textit{et al.} in \cite{HE2014}. They assume the attacker has the ability to manipulate network traffic between a Tor client and an entry node and to collect relevant traffic traces. For the classification part of the website fingerprinting attack, He \textit{et al.} mainly focus on features such as incoming burst volume, incoming burst packet number, total per-direction bandwidth, and total per-direction number of packets. They use SVM with one-against-rest (a technique used when binary classification algorithms are applied to multi-class problems) as the classifier. The authors of \cite{HE2014} compare their work with that of Panchenko \textit{et al.} \cite{PANCHENKO2011} and state that their methods improved accuracy by 16.5\%.}

Arp \textit{et al.} \cite{ARP2015} introduce another side-channel de-anonymisation attack on Tor called \textit{Torben}. This attack exploits an interplay of the following scenarios. 1. The ability to manipulate a web page to load content from an untrusted origin, and 2. the visibility of the size of requests-response pairs of web traffic, regardless of them being encrypted. The basic idea of this attack is to implant a web page marker into the response from the server, which induces a recognisable traffic pattern that a passive attacker can observe at the user's end. Arp \textit{et al.} discuss two variants of the attack, depending on the type of marker. A \textit{remote marker} can be used for web pages that allow content from other origins such as advertisements, and a \textit{local marker} is an item on a web page into which the attacker directly injects content. Arp \textit{et al.'s} attack assumes a real-world adversary who has access to the traffic between the Tor client and the entry node.

\subsubsection{\ik{\textbf{Discussion}}}

\ik{Out of the four categories we defined in our taxonomy, this category is the one with the highest number of attacks, many of which are recent. Most attacks are based on website fingerprinting for several reasons. First, most WF attacks we have discussed are passive attacks (except for \cite{HE2014} and \cite{YANG2017}). A passive attack is stealthier as the victim or a third party may not notice the attack. Also, it requires fewer resources when compared with an active attack that modifies the traffic. Second, the success rates WF attacks are higher when compared with other attacks. Third, WF attacks allow researchers to experiment with new and trending technologies such as Deep Learning (DL) and Artificial Intelligence (AI). In our opinion, the last two reasons strongly motivate the research community to do further work on WF attacks. When looking at Table \ref{table:Attack summary_2}, it is clear that researchers initially used classical machine learning algorithms such as Naive-Bayes, SVM, and k-NN before moving towards more complex deep learning algorithms such as Neural Networks and Autoencoders. The selection of features used in those algorithms also evolved from basic time, packet, and volume-based features \cite{HERRMANN2009,PANCHENKO2011} to compressed \cite{NASR2017} and automatically extracted features \cite{RIMMER2018}. As the development of DL and AI are constantly evolving, we believe that there will be more effective WF attacks in the future. It should also be noted that entities such as Website Oracles \cite{PULLS2020}, when integrated with WF attacks, may present a significant threat to the Tor network in the future. As research on AI and DL is also focused on reducing computational overhead and resource requirements, the size of the Tor network may not provide sufficient anonymity against such advanced attacks. Table \ref{table:Attack summary_2} also shows novelty claims of WF attacks for further reading.}

\ik{Although WF attacks are one of the main types of side-channel attacks, in this paper we have also discussed many other side-channel attacks. Even at the very early stages of Tor, researchers have been discussing adversaries with access to Internet Exchange Points (IXPs), and other AS-level adversaries \cite{MURDOCH2007}. According to a study by Johnson \textit{et al.} \cite{JOHNSON2013}, these are very realistic adversaries. Sun \textit{et al.} even execute a Raptor attack \cite{SUN2015} in the real world with the collaboration of AS operators. These highly resourceful adversaries can easily gain access to Tor traffic flows if they are combined with more advanced correlation techniques such as the ones proposed in \cite{NASR2018} and \cite{PALMIERI2019}. Techniques based on neural networks \cite{NASR2018} and wavelet multi-resolution analysis \cite{PALMIERI2019} are good examples of modern technologies being adopted in Tor research. Another important thing we noticed among the list of attacks in this category is the application of techniques that are more commonly used in different research areas. Profile Hidden Markov Model (PHMM) used in bioinformatics \cite{ZHUO2018}, compressed sensing \cite{NASR2017}, and wavelet multi-resolution analysis \cite{PALMIERI2019} used in signal processing are some examples. These works can encourage more multi-disciplinary research, subsequently opening new avenues in Tor research. In addition to the passive side-channel attacks we have discussed so far, Yang \textit{et al.} \cite{YANG2018} focus on de-anonymising Tor mobile users. As the new Tor Browser for Android has been recently developed \cite{FINKEL2018}, we expect more attention on the anonymity of Tor mobile users.}

\ik{We also discuss a few attacks that we categorise under active side-channel attacks. He \textit{et al.'s} \cite{HE2014} active WF inspired the later work of Yang \textit{et al.} \cite{YANG2017} and considering the popularity of WF attacks in the research community, it is reasonable to expect more improved active WF attacks in the future. However, active attacks are less stealthy than passive WF attacks. Although it is extremely difficult to modify traffic without directly compromising a component of the Tor circuit, the attacks proposed by Gilad \textit{et al.} \cite{GILAD2012} and Arp \textit{et al.} \cite{ARP2015} are interesting. Gilad \textit{et al.} exploit a congestion event while Arp \textit{et al.} use web page markers to induce identifiable traffic patterns. We note that most active attacks that use side-channels also need to compromise one of the Tor circuit components and fall under the Hybrid category.}


\subsection{Hybrid}

If a mix of Tor network components from the previous categories is used for an attack, that attack is categorised under the \textit{Hybrid} category. For example, such an attack requires a combination of a Tor node, a server, a client, and a side-channel.

\subsubsection{\textbf{Passive Attacks}}

\begin{figure}[h]
\centering
\includegraphics[width=8cm]{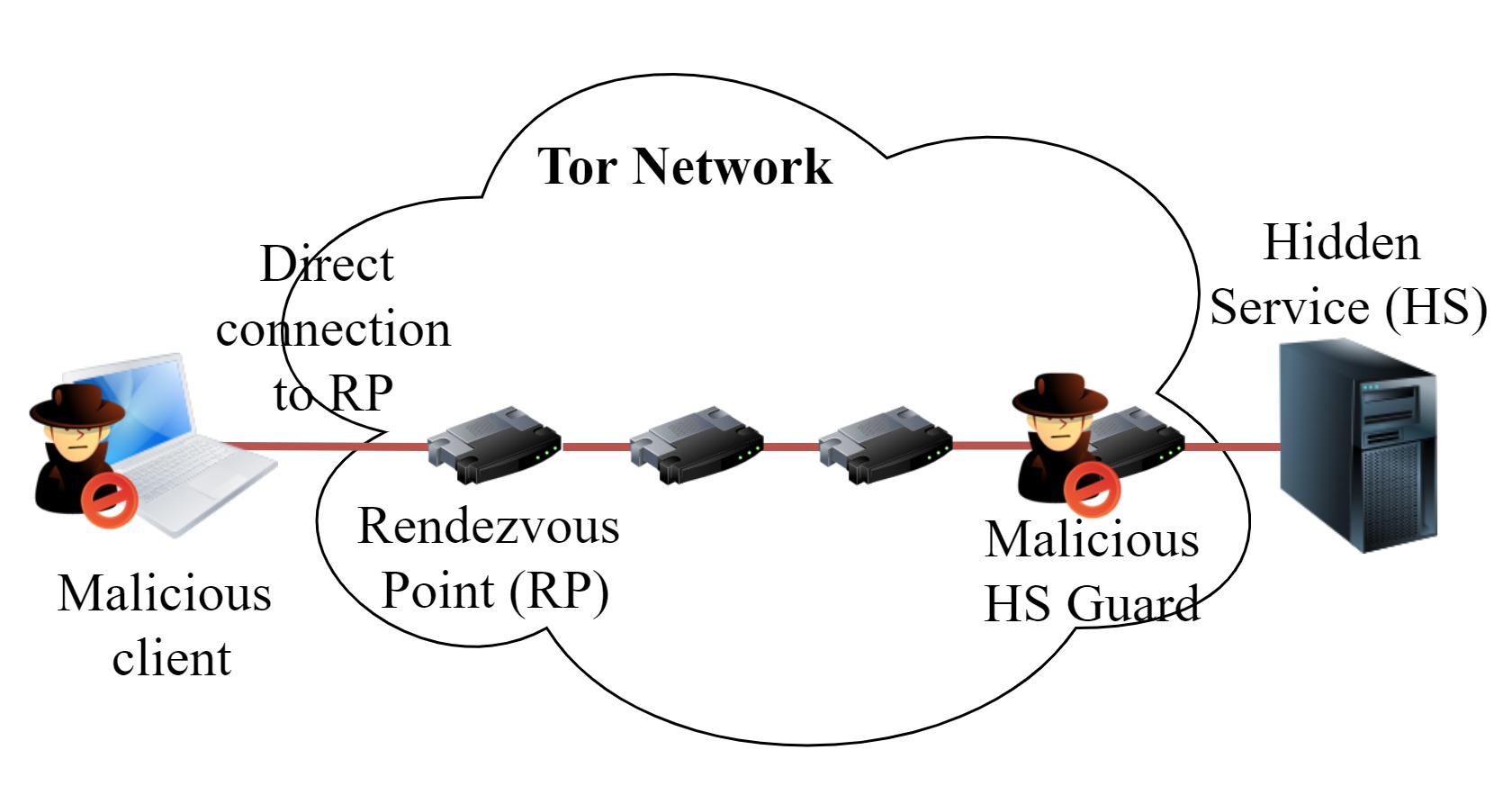}
\caption{Client Scenario in {\O}verlier \textit{et al.}'s HS attack }
\label{fig:overlier attack}
\end{figure}

In 2006, {\O}verlier \textit{et al.} \cite{OVERLIER2006} presented the first known attack on Tor's HSs. In their attack scenario, an attacker-controlled client directly connected to the RP to reduce the latency between the client and the HS. In addition, {\O}verlier \textit{et al.} controlled a \textit{middleman} Tor node, which advertised false up-time and bandwidth, expecting to be selected as a part of the circuit between the HS and the RP. Figure \ref{fig:overlier attack} shows this attack scenario. A traffic pattern was generated from the client expecting to be observed by the malicious node. This technique was used to determine whether the node was a part of the HS circuit. The match confirmation was done by basic packet counting using timing information and the direction of traffic. The next issue faced by the authors of \cite{OVERLIER2006} was to identify the position of the node in the circuit. As the client was aware of the RP's IP address, it could determine when the malicious node was closest to the RP. When this happened, the circuit was torn down, and a new connection was forced in the next attempt. If both IP addresses connected to the attacker node were unknown, follow-up attacks were suggested to determine its position.
1. Service Location Attack - HS could be hosted either on a node in the Tor network (server scenario) or an external client using the Tor network (client scenario). This approach was based on DSs having a public list of all server nodes in the network. If an IP address connected to the attacker node was not available in the public list of nodes, then it had to be the IP address of a HS hosted in an external client (client scenario). 
2. Predecessor attack - The basic concepts of this type of attack \cite{WRIGHT2004} were initially discussed under \textit{Entry and Exit router passive attacks}. By collecting the IP address statistics, the IP address of the HS could be deduced via traffic pattern matches that were found while communicating with the HS.
3. The distance attack - This approach calculated the round trip time of a node's traffic to determine whether it was closer to the HS or not.
Moreover, {\O}verlier \textit{et al.} pointed out that if the attacker owned the RP in addition to the other malicious Tor node, the speed and accuracy of the attack could be increased. The reason was that if the attacker-controlled Tor node was selected as the middle node of the circuit, it could be easily identified as the RP knew the IP address of the node next to it.

Bauer \textit{et al.} \cite{BAUER2010} investigate the benefits and drawbacks of two-hop and three-hop paths in Tor, based on security and performance perspectives. Their paper describes an attack based on \textit{Adaptive Surveillance} in which the objective is to find the identity of the entry guard, assuming that the Law Enforcement Agencies (LEAs) or other powerful adversaries can \textit{adaptively} demand network logs from the entry guard's network. Bauer \textit{et al.} experiment with three-hop scenarios where the attacker controls exit and middle routers to identify the entry guard.

Mittal \textit{et al.} \cite{MITTAL2011} present two types of attacks, one to identify the Tor circuit's guard relays and another one to link two streams multiplexed over the same circuit using \textit{Throughput fingerprinting}.
For the first attack, the attacker should have the ability to observe the throughput of the target flow. This requirement can be achieved by using a compromised exit relay, a web server or an ISP. In this type of attack, one-hop circuits are developed through suspected Tor relays and probed. The attacker does not need to alter the traffic, and the probing can be conducted from a suitable vantage point. If the throughput of the target flow and the probe flow correlate highly, it can be assumed that the experimenting node is a part of the target flow. Mittal \textit{et al.} demonstrate that by doing this for several of a client's circuits and observing the frequency of Tor nodes that are discovered as part of the flow, it can be assumed that these nodes are the circuit's guard nodes as guard nodes have a very high probability of being in a client's circuits. This concept is further used to identify HSs hosted on Tor relays as that relay will have the highest frequency of being part of the circuit.

\subsubsection{\textbf{Active Attacks}}

One of the earliest de-anonymisation attacks against the Tor network was published by Murdoch \textit{et al.} \cite{MURDOCH2005} in 2005. This traffic analysis attack considers that a Tor node's load affects the latency of all connections through that node. The attack uses a corrupt server to send unique traffic patterns, consisting of sequences of short data bursts. The attacker also controls a Tor node, which creates a connection through a targeted node. Following this, the attacker sends probe traffic through the established connection to identify the latency and detect traffic signals passing through the node. This way, the attacker can identify a target circuit's nodes. However, the attacker must access the victim's entry guard to identify the actual originator of the connection. Murdoch \textit{et al.} mention that a simple strategy to use cover traffic does not prevent this type of attack, as the attack depends on indirect measurements of stream traffic. Furthermore, they suggest a few variants of their attack as follows.
1. If the corrupt server cannot significantly modulate the traffic, the modulated probe traffic can be sent to the victim's Tor node in a loop and can detect the effect upon requests sent by the targeted sender.
2. If the attacker cannot modify the traffic but can observe the link, a known traffic pattern on the server can be used for this attack. 
3. If the attacker cannot modulate or observe link traffic, then he/she may resort to observing response times on the server.
4. A DOS attack can be executed on the destination server and can observe the decreased load upon the Tor nodes.
Additionally, the authors of \cite{MURDOCH2005} propose a \textit{linkability attack}, which tries to determine whether two streams coming out of a node belong to the same initiator or not.

Hopper \textit{et al.} \cite{HOPPER2010} present two attacks based on network latency: a de-anonymisation attack and a linkability attack. The de-anonymisation attack is carried out using a malicious client, server, and a Tor node. The first step of the attack is similar to the attack presented by Murdoch \textit{et al.} \cite{MURDOCH2005} wherein the malicious Tor node and the server collude to reveal the nodes in the target circuit. Following this, the attacker estimates Round Trip Times (RTT) for several candidate victims. Afterwards, the malicious client connects to the server using the same set of nodes and checks its RTT to compare it with those of candidate victims.
The authors also propose a linkability attack using two colluding servers and calculating the RTT. 

\begin{figure}[h]
\centering
\includegraphics[width=8cm]{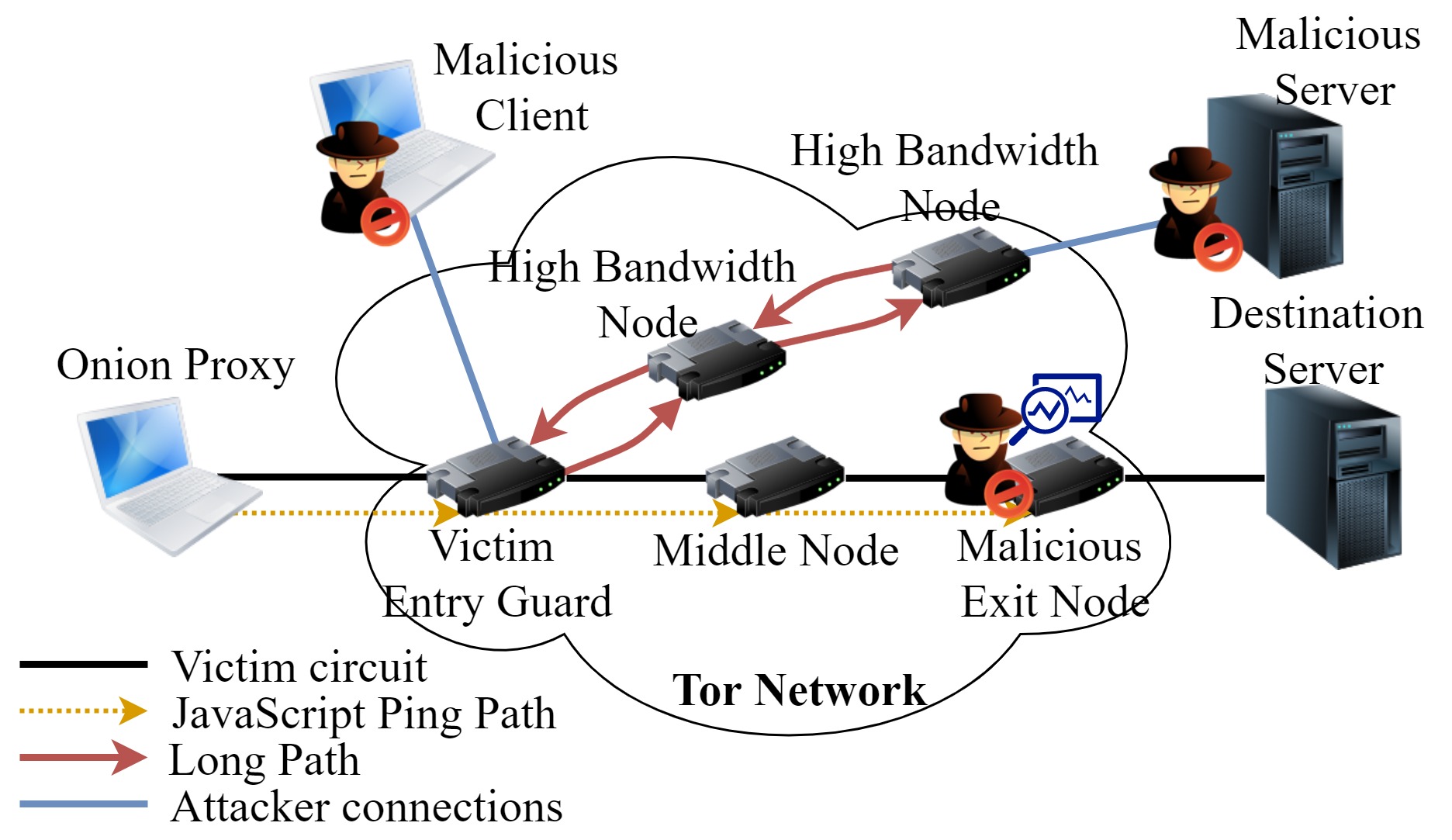}
\caption{Attack Scenario of Evans \textit{et al.} }
\label{fig:evans attack}
\end{figure}

In 2009, Evans \textit{et al.} \cite{EVANS2009} published a paper arguing that the attack presented by Murdoch \textit{et al.} in 2005 \cite{MURDOCH2005} was not practical due to the growth of the Tor network. Evans \textit{et al.} present a new solution by combining the original attack with a novel bandwidth amplification attack. In this scenario, the adversary has to control an exit node, a client, and a server to execute the attack. Firstly, the attacker injects a JavaScript code into an HTML response. This JavaScript code sends periodical HTTP requests, including the time stamp, to which the exit node returns an empty response. Then the malicious client and malicious server create a long path (where the circuit is being looped) for each candidate node (and two helper nodes with high bandwidth). They then start transmitting data to initiate a congestion attack on the candidate node. If the candidate node is part of the target circuit, the exit router observes a significant delay pattern matching the power of the attack. Figure \ref{fig:evans attack} shows the attack scenario for this attack.
It should also be noted that in 2013, Chan-Tin \textit{et al.} \cite{CHAN-TIN2013} revisited the Murdoch \textit{et al.} attack \cite{MURDOCH2005} and demonstrated that it was still possible to execute regardless of Evans \textit{et al.'s} claims. The attack by Chan-Tin \textit{et al.} is very similar to that of Murdoch \textit{et al.}, with a few modifications.

Chakravarty \textit{et al.} \cite{CHAKRAVARTY2010} present an attack by estimating the bandwidth of Tor nodes by using \textit{available bandwidth} estimation tools. They used \textit{LinkWidth}, a tool that can estimate the available bandwidth and capacity bandwidth on a path \cite{CHAKRAVARTY2008}. This type of attack requires a malicious server to inject bandwidth fluctuations into the circuit. Following this, all Tor nodes are probed repeatedly to detect these bandwidth fluctuations using LinkWidth. However, this process only allows the attacker to identify the Tor nodes in the target circuit. In order to find out the identity of the Tor client, the attacker has to monitor the fluctuations on the link between the client and the entry node, an action that would only be practical to an AS-level adversary or an ISP. This process can also be used to identify HSs using a malicious client instead of a server to induce bandwidth fluctuations.

Ling \textit{et al.} \cite{LING2013} discuss a type of attack that can be used to de-anonymise HSs. For this, the attacker has to have control over a Tor client, an RP, some entry routers, and a central server. The attack is carried out in three steps.
1. Identifying the HS, presumably by trying to connect it to one of the compromised entry nodes. A unique combination of cells of different types is used for this.
2. Confirming the HS by sending a modified cell from the RP, which destroys the circuit.
3. Recording the details about timing and IP addresses from the RP and entry routers in the attacker-controlled central server, which is finally correlated to conclude the HS's IP address.
Ling \textit{et al.} claim that their attack's true positive rate is 100\% but mention two complications of this attack.
1. The typical expiry date of entry guards is 30-60 days, and if no compromised routers are selected as HS entry guards, this approach fails.
2. If the operator selects trusted routers or bridges as entry guards, the attacker has to resort to an additional step to find out what these routers are and compromise them before executing the original attack.

\begin{table*}[t]
  \footnotesize
  \centering
  \caption{Summary of Hidden Service attacks}
  \label{table:Attack summary_3}
  \begin{threeparttable}
    \begin{tabular}{ |l|l|l|l|l|l|l| } 
      \hline
      \textbf{Publication} & \textbf{Year} & \textbf{Components used} & \textbf{Active/Passive} & \textbf{Comments}
      \\
      \hline
      {\O}verlier \textit{et al.} \cite{OVERLIER2006} & 2006 & HS Entry and OP & Passive & First attack on HSs
      \\
      \hline
      Murodch \cite{MURDOCH2006} & 2006 & Onion proxy & Passive & Use changes in clock skew
      \\
      \hline
      Zander \textit{et al.} \cite{ZANDER2008} & 2008 & Onion proxy & Passive & Reduce quantisation error in Murdoch 2006 attack
      \\
      \hline
      Chakravarty \textit{et al.} \cite{CHAKRAVARTY2010} & 2010 & OP and LinkWidth & Active & Use \textit{available bandwidth} estimation tools
      \\
      \hline
      Mittal \textit{et al.} \cite{MITTAL2011} & 2011 & Onion proxy & Passive & Can only be used for HSs hosted in Tor nodes
      \\
      \hline
      Ling \textit{et al.} \cite{LING2013} & 2013 & HS Entry, OP and RP & Active & Protocol-level discovery approach
      \\
      \hline
      Biryukov \textit{et al.} \cite{BIRYUKOV2013} & 2013 & HS Entry, OP and RP  & Active & Use a special padding cell technique in Tor
      \\
      \hline
      Chakravarty \textit{et al.} \cite{CHAKRAVARTY2014} & 2014 & HS Entry and OP & Active & Use NetFlow traffic records
      \\
      \hline
      Matic \textit{et al.} \cite{MATIC2015} & 2015 & Onion proxy & Passive & Use location leaks in HS's content and configuration
      \\
      \hline
      Kwon \textit{et al.} \cite{KWON2015} & 2015 & HS Entry node & Passive & Circuit fingerprinting attack for HS
      \\
      \hline
      Hayes \textit{et al.} \cite{HAYES2016} & 2016 & ISP/Network admin & Passive & Propose k-fingerprinting against both normal services and HSs
      \\
      \hline
      Panchenko \textit{et al.} \cite{PANCHENKO2017} & 2017 & ISP/Network admin & Passive & Evaluate the scalability of fingerprinting attacks on HSs
      \\
      \hline
      Iacovazzi \textit{et al.} \cite{IACOVAZZI2018} & 2018 & HS Entry and OP & Active & First method to embed watermarks at destination and detect at source
      \\
      \hline
      Iacovazzi \textit{et al.} \cite{IACOVAZZI2019} & 2019 & HS Entry and OP & Active & Exploits a weakness in Tor's congestion control mechanism
      \\
      \hline
    \end{tabular}
   \end{threeparttable}
\end{table*}

Gedds \textit{et al.} \cite{GEDDS2013} introduce a set of attacks, known as \textit{Induced Throttling Attacks}. These attacks assume a scenario in which the attacker controls an exit node and some middle nodes to identify candidate entry guards. 
One attack exploits a feature of Tor's congestion control algorithm. These algorithms send ``backward control cells" to request data from edge nodes. When the nodes do not receive these cells, they stop sending data until the next cell is received. A malicious exit node can use this mechanism to create congestion artificially and induce specific patterns in a circuit. This phenomenon affects all circuits going through the relevant node. The attacker then uses malicious middle nodes to create one-hop probe circuits through suspicious entry guards to identify these patterns by measuring the throughput and congestion on these circuits at regular intervals.
The second attack also utilises the fact that the throttle rate of a circuit depends on the number of \textit{client-to-guard} connections. To execute this attack, the adversary only needs a malicious client and an exit node. The adversary can create multiple connections to a guard node using the client and toggle its throttling rate while the exit node tries to identify the pattern. However, as this attack can only be used to identify the entry guard, additional steps are required to de-anonymise the sender.

Sulaiman \textit{et al.} \cite{SULAIMAN2013} describe an attack using unpopular ports in Tor. Many volunteers restrict certain ports from use in their Tor nodes for reasons like bandwidth greediness, associated spamming and de-anonymisation risks. Some of these ports include SMTP (25), NNTP (119), NNTPS (563), and many P2P ports. The attack explained in \cite{SULAIMAN2013} happens in several steps. The attacker first compromises a web server and then injects malicious entry and exit routers into the network, each advertising a high bandwidth. These exit routers allow targeted unpopular ports. When a client connects to the compromised web server through a \textit{popular} port, it sends back a hidden script with the requested web page. When the client executes the hidden script, the script forces the client to create a connection through Tor using the targeted \textit{unpopular} port. If a circuit is established via two malicious nodes, the attacker can use traffic analysis to de-anonymise the user. 

Chakravarty \textit{et al.} \cite{CHAKRAVARTY2014} present an attack using NetFlow records. They mention two attack scenarios; one that de-anonymises the client and one that de-anonymises a HS.
For the first attack, the adversary must control a malicious server and entry nodes. The attacker then injects traffic patterns that are identified by the entry guards.
In the second attack, a malicious Tor client injects a traffic pattern that can be identified by an adversary-controlled HS entry node. The flow records in these scenarios are used to calculate a correlation coefficient to link the two endpoints.

Biryukov \textit{et al.} \cite{BIRYUKOV2013} suggest an attack to de-anonymise HSs. In this type of attack, the attacker requires a malicious client, an RP, and some guard nodes. When a client tries to connect to the HS, it sends a Rendezvous Cookie (RC) and the RP's address to an introduction point of the HS (refer Figure \ref{fig:Hidden server connection}). Following this, the introduction point communicates these details to the HS. The HS sends a RELAY\_COMMAND\_RENDEZVOUS1 cell containing the RC to the RP to build the circuit between RP and the HS. Upon receiving this, the malicious RP sends 50 padding cells back to the HSs (which are discarded by the HS), followed by a DESTROY cell. If there is an attacker-controlled guard node in this circuit, it is identified as follows.
1. Whenever the attacker node receives a DESTROY cell, it checks whether the DESTROY cell was received just after the RELAY\_COMMAND\_RENDEZVOUS1 cell. 
2. Following this, it checks whether the number of cells forwarded from the HS is 3 (2 RELAY\_COMMAND\_EXTEND cells + 1 RENDEZVOUS1 cell) and transmitted to the HS is 53 (50 Padding Cells + 2 RELAY\_COMMAND\_EXTENDED cells + 1 DESTROY cell). 
If both of these conditions are met, then the attacker node is selected as a guard node. As the guard node knows the IP address of the HS, it can be de-anonymised. The same concept can be used to identify when the attacker node is selected as the middle node of the HS-RP circuit when the numbers of forwarded cells are 2 and 52 (without 1 RELAY\_COMMAND\_EXTEND cell and RELAY\_COMMAND\_EXTENDED cell, respectively). In this scenario, the attacker can identify the guard nodes and try to compromise or block them.

\ik{New technologies based on ultrasound have recently emerged. These can be used for activities such as device pairing, proximity detection, and cross-device tracking\footnote{\ik{This is a technique generally used for advertising to users and/or tracking them across multiple devices.}}. Inaudible ultrasound beacons that can be emitted by regular speakers and captured by microphones are used in these technologies. When using ultrasound cross-device tracking (uXDT) in advertising campaigns, the advertising client (a company or an individual) first needs to set up an advertising campaign and send ads to an uXDT provider. A uXTD provider is a company that can provide the required infrastructure to advertising clients. After agreeing to provide services to the client, the uXTD provider generates a unique ultrasound beacon associated with the client's campaign and sends it to a content provider (e.g. social media, websites, TV) along with the client's ads. When a user views one of those ads, the respective beacon is emitted from the user's device (e.g., laptop speaker, TV) and is captured by a uXDT enabled device (e.g. smartphone). The uXDT enabled device then sends some information related to this incident to the uXDT provider, which can be used to identify the user's interests. Mavroudis \textit{et al.} \cite{MAVROUDIS2017} have used this ultrasound-based technology to execute a de-anonymisation attack on Tor. The attacker requires an ultrasound enabled device and access to a Tor exit node or a hidden service to execute this attack. Similar to the scenario explained above, the attacker first needs to create a campaign with a uXDT provider and to obtain an ultrasound beacon. Then, the attacker can inject a code snippet that causes the emission of the ultrasound beacon into the traffic using a compromised exit node or a server. Once the user's browser receives this code snippet, it will start emitting the beacon through a speaker. This beacon will be captured by a uXDT enabled device (e.g. smartphone), and relevant details will be sent to the uXDT service provider. An attacker such as an LEA can obtain this information from the service providers via legal procedures.}

Iacovazzi \textit{et al.} \cite{IACOVAZZI2018} introduce an inverse flow watermarking attack called \textit{INFLOW} to de-anonymise HSs. They argue that in previous watermarking attacks, the attack was only effective in tracking watermarks from source to destination, as a watermark only travels in the direction of the traffic flow. Iacovazzi \textit{et al.} claim that INFLOW is the first technique that can insert a watermark at the destination, which is detectable at the source. It exploits Tor's congestion management, which stops sending messages until an ACK for the previous message is received. When the user downloads large amounts of content from a server, there are increased traffic flows from the server to the client, whereas only a few packets and ACKs are transmitted from the client to the server. Therefore, sometimes this traffic from the client to the server may not be enough to ensure that watermarks are embedded. However, using this concept, a modified client can drop bursts of ACKs and prompt traffic patterns from the server, which can be identified by an attacker-controlled HS guard node.

Iacovazzi \textit{et al.} \cite{IACOVAZZI2019} present a more recent active traffic analysis attack called the \textit{Duster Attack}. In this attack, the objective is to de-anonymise HSs. A malicious client and a set of guard relays are used for the attack. Firstly, the attacker selects a HS and starts downloading content from it after establishing a connection. During the data transfer, a watermark is injected into the traffic by the Tor client. If a malicious guard detects this watermark, it will record the IP address of the circuit endpoint and cancel the watermark. Tor uses a technique based on SENDME cells for congestion control, which can be exploited to embed the signal into the Tor traffic in this type of attack. Iacovazzi \textit{et al.} experiment on the live Tor network, and claim that this attack possesses the following properties, lacking in previous active attacks in the literature: 1. it works on both standard and rendezvous circuits, 2. it is hidden from the target endpoint, 3. it has a small overhead and does not affect network performance, 4. it exploits vulnerabilities in Tor's congestion control mechanism, and 5. it works on Tor versions up to 2019. 


\subsubsection{\ik{\textbf{Discussion}}}

\ik{We note that many de-anonymisation attacks on HS fall in the hybrid category. Table \ref{table:Attack summary_3} provides a summary of all the HS attacks discussed in this paper. By the end of 2020, there were more than 170000 unique .onion addresses, which shows that the HS space has grown by about a factor of five since 2015 \cite{TORMETRICS}. With this growth in the Tor network, it is reasonable to assume that some of the hidden service attacks are either completely unrealistic or difficult to execute in today’s Tor network. For example, attacks that need to probe Tor nodes \cite{CHAKRAVARTY2010} or machines \cite{MURDOCH2006,ZANDER2008} to de-anonymise HSs may find their execution difficult due to the current size of the Tor network. From Table \ref{table:Attack summary_3}, we can also see that the de-anonymisation of HSs has been an active area of research over the years. As there have been real-world attacks on HSs \cite{EUROPOL}, where several HSs were identified, it is imperative to continue research on HS de-anonymisation. We can also note that some works have tried to use WF on HSs, a concept further analysed by Overdorf \textit{et al.} \cite{OVERDORF2017}. Overdorf \textit{et al.} conclude that WF attacks are successful on HSs and certain HSs are more likely to be de-anonymised based on factors such as the size and the dynamics of the HS. The more recent watermark attacks on HSs \cite{IACOVAZZI2018,IACOVAZZI2019} are reported to be effective. These attacks exploit Tor’s congestion control mechanism and, having been tested against more recent versions of Tor, present a major threat to the anonymity of HSs.}

\ik{To summarise, in general, network growth and enhanced security measures have invalidated some of the legacy de-anonymisation attacks on Tor. However, the evolution of these attacks can still overwhelm Tor's current network defences. As a case in point, 4 years after Murdoch \textit{et al.'s} \cite{MURDOCH2005} attack in 2005, Evans \textit{et al.} \cite{EVANS2009} claimed that the attack reported in \cite{MURDOCH2005} was no longer practical due to the network’s growth. However, in 2013 Chan-Tin \textit{et al.} \cite{CHAN-TIN2013}] revisited the attack described in \cite{MURDOCH2005} and claimed that it was still practical, with a few modifications. This also indicates why a performance comparison with previously reported attacks is important. Additionally, the attack based on ultrasound cross-device tracking (uXDT) \cite{MAVROUDIS2017} is another good example of new technologies opening doors to novel and creative attacks. Throughout this paper, we have discussed Internet Service Providers, Network Administrators, Internet Exchange Points, and AS operators as realistic adversaries. The attacks \cite{MAVROUDIS2017,YANG2018} show that the scope of potential adversaries is also widening. For example, the uXDT attack can easily be executed by a uXDT service provider. Therefore, it is important to keep track of the evolution of existing attacks and the emergence of new attack vectors.}


\section{\ik{Security against De-anonymisation}}

\ik{In this section, we provide insights into some of the main security improvements and changes made to the Tor protocol to protect it against de-anonymisation attacks. When trying to evaluate how to mitigate attacks against Tor, we noticed that many papers over the years suggest various countermeasures and security improvements, warranting a separate comprehensive survey. Additionally, the Tor Project maintains a blog (https://blog.torproject.org/), which also provides some of the developments and fixes that have been implemented to improve Tor’s security. To keep this section manageable, we decided to limit the discussion to articles from the Tor blog covering de-anonymisation attacks. Hence, we do not consider any specific work related to countermeasures or security improvements in Tor literature unless they are associated with a Tor blog article. The articles in the Tor blog have been written by authors actively working with the Tor development team and cover realistic scenarios.} A three-part article series \cite{MATHEWSON2012_1,MATHEWSON2012_2,MURDOCH2012_3} written in 2012, presents major changes in Tor since its initial design paper in 2004 \cite{DINGLEDINE2004}. We will first discuss the security-related changes in these articles and then move on to some of the later updates.

\subsection{Security Improvements in the Directory System}
In the initial Tor versions, every router in the network generated a \textit{router descriptor} that was signed and uploaded onto one of the DSs \cite{MATHEWSON2012_1}. Every DS created a signed list of its descriptors and sent them to the clients at their request. Among the many issues of the above mechanism were a few security concerns as well. 1. There was no distributed trust, and each DS was trusted individually. Hence, a compromised DS could be used to execute an attack on all its clients. 2. The directory content was not encrypted and easy to eavesdrop on. 3. Disagreeing DSs could divide client knowledge, allowing an attacker to categorise clients based on their recent connections to DSs. There were a few changes in DSs along the way, including assigning flags (such as \textit{fast, stable, good guard nodes}) to nodes before the introduction of a directory voting system. Under the directory voting system, the DSs would share periodic vote documents and produce a \textit{consensus document}, and every DS would sign it. As a result, the clients only need to download one document and make sure a majority of known DSs signed it. These changes helped to address some security issues that were in the original design. 

\subsection{Introducing Guard Nodes}
In Section II, we briefly introduced guard nodes and their purpose. Several of the early de-anonymisation attacks were based on having an adversary-controlled entry node. Based on the recommendations of {\O}verlier \textit{et al.} \cite{OVERLIER2006}, Tor implemented the \textit{Guard Node} feature to reduce the probability of circuits being compromised \cite{MATHEWSON2012_2}. Ordinary Tor nodes are now assigned guard flags based on several features such as bandwidth and uptime. Once a Tor client selects a set of guard nodes, it will keep them for 30-60 days. This situation greatly reduces the chances of an adversary compromising circuits by introducing new Tor nodes and expecting them to be selected as entry nodes. On the other hand, if an adversary-controlled node is selected as a guard node, the adversary has a significant chance of de-anonymising the user, as that node will repeatedly be used for many circuits for a considerable duration.

\subsection{Introducing Bandwidth Authorities}
The nodes of a circuit were uniformly picked at random in the earliest versions of Tor. However, this created many bandwidth bottlenecks, hugely impacting Tor's performance. As a result, the Tor protocol was changed to select nodes proportionally to the node's bandwidth and based on its capabilities (e.g. entry guard, exit node). This feature increased the chances of attackers compromising more circuits by claiming high bandwidths for the nodes under their control. Initially, a maximum bandwidth limit was imposed to minimise the impact of this, but later a set of \textit{Bandwidth Authorities} were assigned to measure and vote on the observed node bandwidth. These measured values were published in the \textit{consensus} document, preventing the previous loophole. Also, honest node operators were allowed to declare their nodes under the same \textit{family}, to stop the client from selecting two nodes from the same family in the client's circuit. These node families prevented node operators from unintentionally having their nodes selected as both entry and exit nodes of a circuit.

\subsection{Mitigating Linkability Attacks}
We mentioned a couple of attacks related to linking Tor streams in Section V (e.g. the bad apple attack \cite{LEBLOND2011, MITTAL2011}. Due to the computational and bandwidth overhead to the network when creating new circuits, Tor clients try to reuse circuits, sending multiple TCP streams through them. The issue with this scenario is that if one stream leaks information to de-anonymise the user, a compromised exit node may be able to de-anonymise other streams in that circuit as well \cite{LEBLOND2011}. Tor limits the circuit usage time to ten minutes before switching to a new circuit to mitigate this risk. The Tor user also has the ability to create new circuits for new streams and configure Tor to isolate streams based on the destination IP/port. By default, Tor separates streams from different clients, different SOCKS ports on a Tor client or SOCKS connections with distinct authentication credentials.

\subsection{A Defence Against Website Fingerprinting}
We discussed several WF attacks on Tor over the years. However, it was only following Panchenko \textit{et al.'s} attack \cite{PANCHENKO2011} in 2011 that Tor developers became concerned with securing the Tor network from such attacks. Panchenko \textit{et al.} used new features based on volume, timing and the direction of traffic, which led to the successful outcome of their attack. In an article in 2011 \cite{PERRY2011}, its author mentions an experimental defence deployed against WF attacks. The proposed defence - introduced as \textit{Random pipelining} - was aimed at reducing the information leakage that enabled the extraction of features used by Panchenko \textit{et al.}. Additionally, the author of the article questions the practicality of WF attacks in the live Tor network, a notion that was critically analysed and supported later by the work of Juarez \textit{et al.} \cite{JUAREZ2014}.

\subsection{Security Against the Relay Early Confirmation Attack}
In July 2014, the Tor development team identified a set of malicious relays in the live network which they believed were trying to de-anonymise Tor users \cite{DINGLEDINE2014_RELAY_EARLY}. These relays were in the network for almost five months and were able to obtain \textit{Guard} and \textit{HSDir} flags. The compromised relays were used to execute an attack known as the \textit{relay early confirmation attack}, which was an active attack that exploited a vulnerability in Tor's protocol headers. A new type of cells known as \textit{relay early cells} were introduced in 2008 to prevent the creation of long paths in Tor circuits \cite{EVANS2009}. These cells were used in the above attack, hence giving the attack its name. In this attack, when the onion proxy tries to either publish or retrieve a service descriptor from an attacker-controlled HSDir node, it will insert a signal into the traffic using the vulnerability in relay early cells mentioned above. The attacker-controlled guard nodes are then able to identify this signal at the other end. After finding out about this issue, all the malicious relays were removed from the network. Moreover, a fix for the issue was given with the next version update of the Tor protocol. Additionally, the above incident brought to attention the importance of monitoring bad relays in the Tor network. Therefore, any interested party can now report any suspicious relays (malicious, damaged or misconfigured) to the Tor Project \cite{BADRELAYS}. The development team would then investigate the issue and take necessary actions.


\subsection{Deploying a Padding Scheme}
Following up on the traffic analysis attack suggested by Chakravarty \textit{et al.} \cite{CHAKRAVARTY2014} that makes use of flow records, the Tor development team published a couple of articles explaining that although this type of attack is theoretically possible, it would be hard to execute it practically against the live Tor network \cite{DINGLEDINE2014_NETFLOW, LEWMAN2014_NETFLOW}. They interpreted the 6\% false-positive rate\footnote{In the original paper, this value is 5.5\% although in the article \cite{DINGLEDINE2014_NETFLOW}, this is used as 6\%} as meaning that 6000 out of 100000 flows will look similar, rendering it ineffective when scaled. Furthermore, the authors of the articles asserted that Tor protects its users by encryption of data within the Tor network, authentication of relays and use of signatures to make sure all clients have the same relay information. In another article from 2015 \cite{DINGDELINE2015} written about the circuit fingerprinting attack against HSs by Kwon \textit{et al.} \cite{KWON2015}, the author of the article brings up a similar argument about Kwon \textit{et al.'s} attack not being scaled in reality. However, in response to the approach introduced by Kwon \textit{et al.} to identify traffic between different circuit components, Tor has deployed a padding scheme to disguise the client-side traffic in HS circuit creation \cite{PADDING_SPEC}. 

\begin{figure}[h]
\centering
\includegraphics[width=8cm]{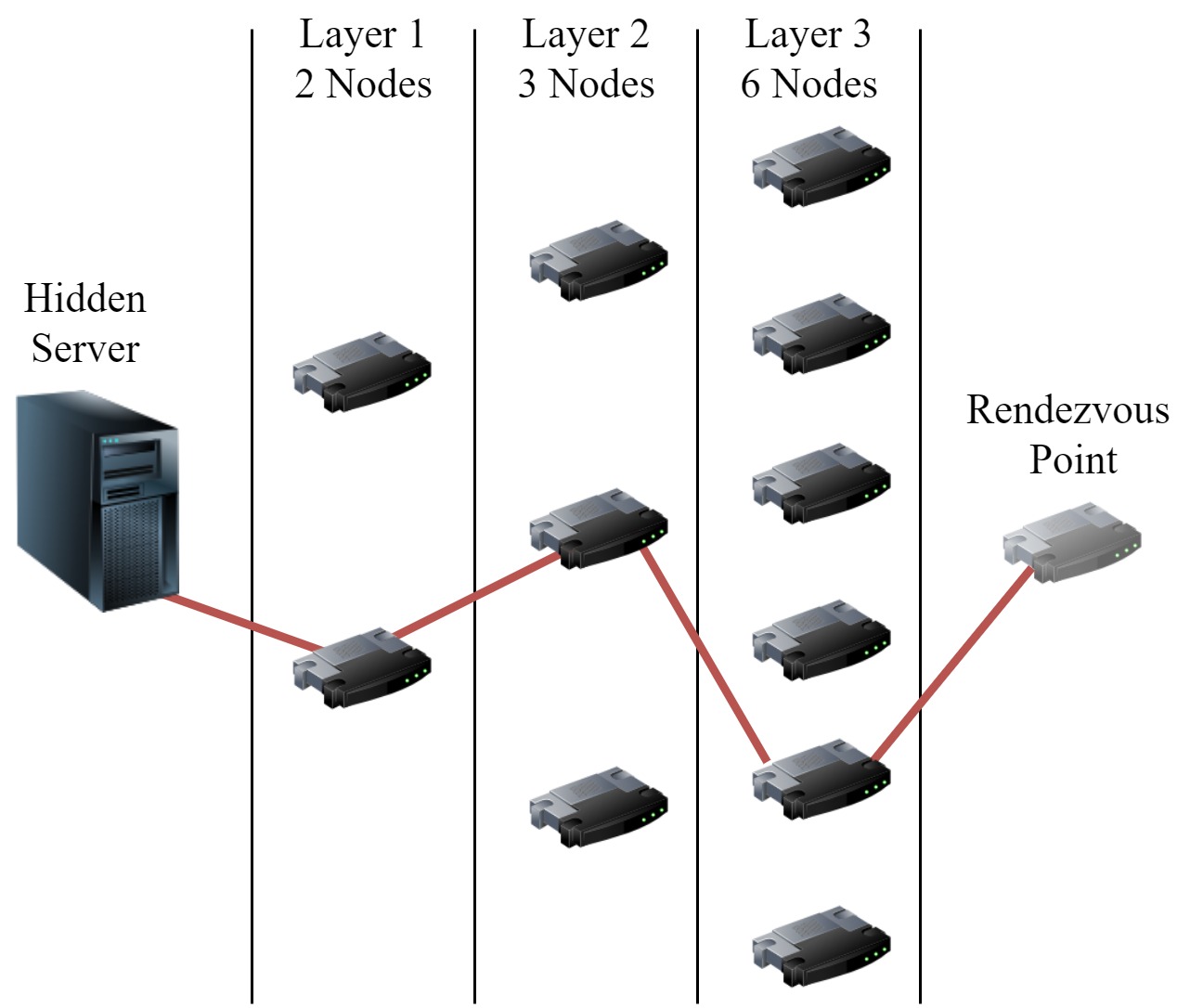}
\caption{Vanguard system: 2-3-6 Topology  }
\label{fig:vanguard}
\end{figure}

\begin{figure*}[h]
\centering
\includegraphics[width=\textwidth]{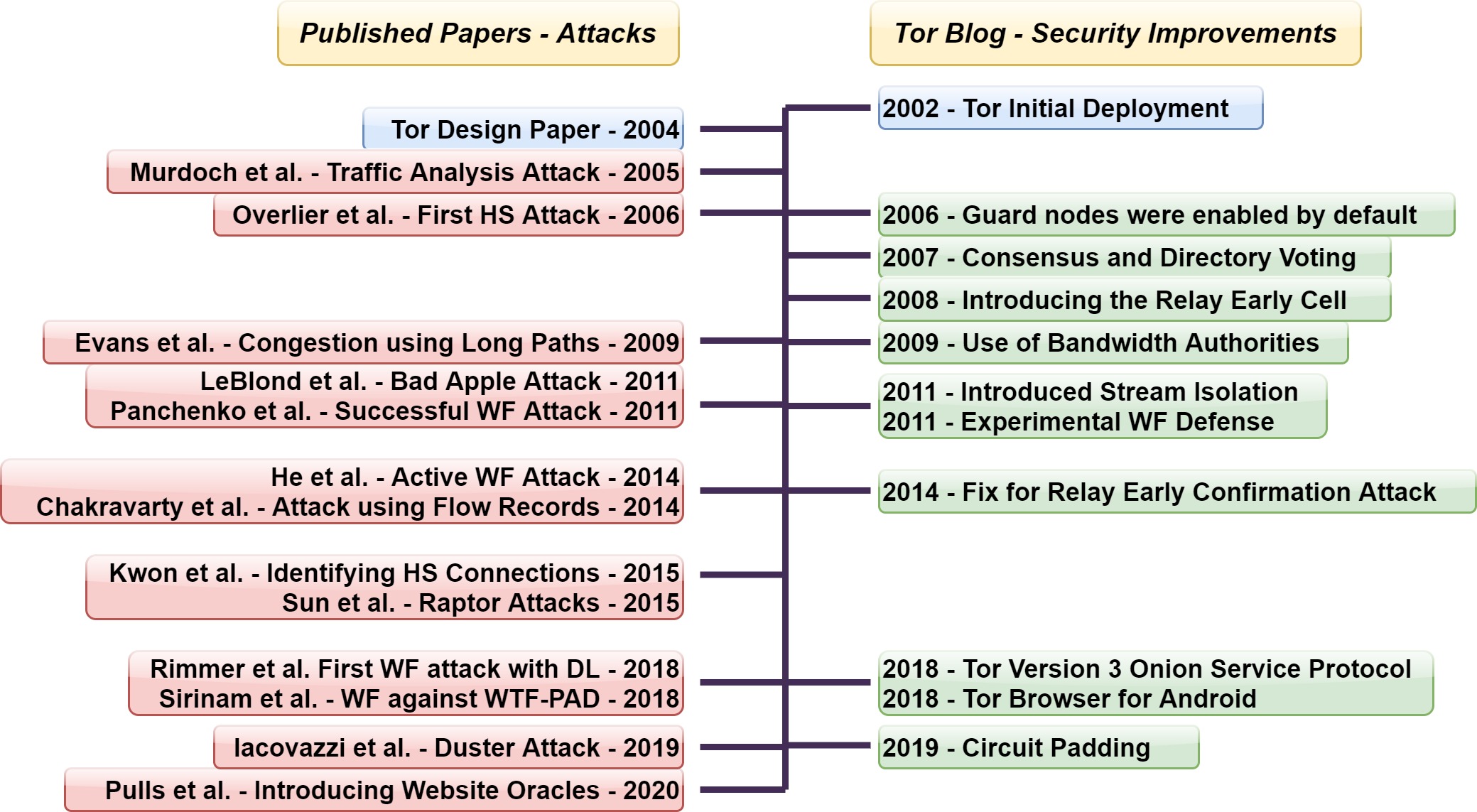}
\caption{Several Security Improvements and Important De-anonymisation Attacks Over the Years}
\label{fig:timeline}
\end{figure*}

\subsection{Introducing the Vanguard System}
There have been real-world attacks against Tor hidden services. In November 2014, 16 European countries, along with the United States Intelligence agencies, brought down several marketplaces hosted as HSs through a coordinated international action called \textit{Operation Onymous} \cite{EUROPOL}. Under this operation, the LEAs took down 410 HSs and arrested 17 people suspected of running the operations. The notorious dark market known as the Silk Road 2.0 \cite{CHRISTIN2013} was also shut down, and its operator was arrested. In response to this massive scale attack, the Tor development team acknowledged that they were unsure why this attack was carried out and how it was carried out \cite{LEWMAN2014_ONYMOUS}. In an article published shortly after the above attack \cite{LEWMAN2014_ONYMOUS}, they discuss possible scenarios that might have enabled this attack. HS operators' failure to use adequate operational security measures, the exploitation of web bugs such as SQL injections or remote file inclusions are some such scenarios. The authors of \cite{LEWMAN2014_ONYMOUS} advised the HS operators to be better informed about HSs' security limitations and ensure their services do not lack adequate memory, processing, and network resources. They also suggested manually selecting known and trusted relays as the guard nodes of the HS.

In 2018, the Tor Project released the first stable version of Tor and the Tor browser for the V3 onion service protocol \cite{KADIANAKIS2018}. According to the article \cite{KADIANAKIS2018}, this protocol version provides more secure onion addresses and improved authentication to HSs. It further provides service enumeration resistance and upgraded cryptography. However, the authors of this article claim that the new upgrade does not address any de-anonymisation attacks against HSs. Also, they state that the highest threat faced by HSs at present are the \textit{Guard Discovery} attacks. As the second and third nodes of the HS-RP circuit are selected from all existing Tor relays in the network, an adversary can force circuits until it is selected as the middle node and can then identify the guard node \cite{BIRYUKOV2013, JANSEN2014}. The guard nodes can then be compromised, attacked or surveilled until the IP address of the HS is obtained. To address this issue, Tor introduces a 3-component add-on to be used with HSs \cite{KADIANAKIS2018}. As shown in Figure \ref{fig:vanguard}, the \textit{Vanguard} component introduces second and third layer guards. In addition, to further increase the anonymity provided by this system, the circuit lengths (e.g. Client - RP, HS - RP) will also be altered. The \textit{Bandguard} checks for bandwidth side-channel attacks and closes circuits more than 24 hours old, and circuits transmitting the maximum threshold of megabytes. The third component of the add-on, known as the \textit{RendGuard}, analyses RPs to check whether they have been overused. This component aims to minimise the use of malicious RPs in potential attacks.

\subsection{Mitigating the Risks of Website Oracles (WO)}
In a follow-up article \cite{PERRY2019} to the WF attack with WO by Pulls \textit{et al.} \cite{PULLS2020}, the Tor development team mentions that they are concerned about the use of low cost, high coverage WOs such as DNS and RTB \cite{JUNWANG2017} to assist attackers. In \cite{PERRY2019}, the authors suggest some precautions that various user groups can take to mitigate the risks of this situation. Users can engage in multiple activities at once with the Tor client, exit relay operators can avoid high-risk public DNS resolvers and stay up to date with Tor releases, and HS operators can use V3 onion services. Additionally, in an article on \textit{Browser Fingerprinting} \cite{LAPERDRIX2019}, a technique that is becoming popular to de-anonymise users, it is explained that Tor has always been concerned with such techniques, and the Tor browser is currently one of the most resilient browsers against browser fingerprinting attacks.

\subsection{\ik{Discussion}}

\ik{In Section V, we discussed several de-anonymisation attacks on Tor. Most of those attacks were experimented in restricted or simulated environments. Therefore, some of them may not be very realistic when applied to the actual Tor network. Moreover, there could be other attacks whose effectiveness has been reduced over time with the growth of the Tor network. Finally, the development team behind Tor has been actively working in the last two decades to improve its performance and security, which might have rendered some of the attacks useless. In this section, our objective was to highlight the evolution of the Tor network, considering several well-known de-anonymisation attacks and security concerns that have been discussed over the years. For this, we used Tor blog articles that we found to be relevant to de-anonymisation attacks. Figure \ref{fig:timeline} shows a summary of most of the security improvements we discussed in this section. It also shows a timeline of those improvements\footnote{\ik{For most improvements, the article we referred to mentions the Tor release version in which the changes were deployed. We cross-referenced these version numbers with the dates in the Tor package archive (https://archive.torproject.org/tor-package-archive/) to find the relevant year.}} along with some relevant publications. We note that most of the security features introduced in the Tor protocol were in response to attacks and findings from the research community. For example, Guard nodes were introduced based on Overlier \textit{et al.'s} \cite{OVERLIER2006} recommendations, relay early cell was introduced to prevent long path creation \cite{EVANS2009}, random pipelining was implemented after Panchenko \textit{et al.'s} WF attack \cite{PANCHENKO2011}, stream isolation was introduced after the Bad Apple attack \cite{LEBLOND2011}, and circuit padding was introduced against WF attacks \cite{KWON2015}.}

\ik{We have referenced articles from the Tor blog for several reasons. First, several articles discuss and evaluate the validity of attacks published in research papers. The articles in references \cite{LEWMAN2014_NETFLOW} and \cite{DINGLEDINE2014_NETFLOW} discuss Chakravarty \textit{et al.'s} traffic correlation attack using netflows \cite{CHAKRAVARTY2014} and show that realistically, the attack in \cite{CHAKRAVARTY2014} is not effective against the live Tor node. \cite{LEWMAN2014_NETFLOW} asserted that attacks that link the entry and exit are very difficult to execute in practice, although they are theoretically possible. The blog article in \cite{PERRY2019} discusses the practicality of a WF attack with WOs \cite{PULLS2020}. While agreeing with the fact that some WOs may present a threat to the Tor network, this article also provides insights about the fact that WF attacks are not very realistic. \cite{PERRY2019} discusses how circuit padding is a strong defence against WF, except for some recent attacks using Deep Learning \cite{SIRINAM2018}. These types of discussions are helpful to researchers as they evaluate the current state of attacks. Second, these articles provide information about actual attacks that have happened \cite{DINGLEDINE2014_RELAY_EARLY,LEWMAN2014_ONYMOUS} and the precautions taken against them. It is important to follow up on such realistic attacks to identify further vulnerabilities and to conduct more effective research.}


\section{CONCLUSIONS AND FUTURE WORK}

In this paper, we first classified Tor attacks into four main categories based on the objective of the attack and explained those categories with examples. Following this, we elaborated on de-anonymisation attacks with a taxonomy based on the components used for attack execution. \ik{Under the classification of de-anonymisation attacks, we provided a corpus of attacks published in the literature, giving brief descriptions on how the attacks are executed and how de-anonymisation is achieved. We then discussed each attack category's significant features while giving some insights into future work and highlighting unique attacks}. We also provided insights into the evolution of these attacks over the years. Finally, we discussed several security-related issues in Tor using the information on articles written by the Tor development team. 

We noticed a few important features while completing this work. 
1. Most of the earlier de-anonymisation attacks focus on compromising network components of the Tor circuit. The main reason for this was the low number of relays in the Tor network when they were published. However, with Tor's increasing popularity, the number of voluntary relays has increased, and the practicality of the attacks that can be executed by compromising a small set of Tor relays has decreased. Therefore, recent attacks assume passive adversaries that can observe the traffic at the source and destination links. 
2. Techniques and concepts from other research domains have inspired Tor researchers to introduce novel attack schemes for Tor \cite{NASR2017,ZHUO2018,MAVROUDIS2017}. \ik{This type of multi-disciplinary research will allow researchers to design more creative and robust attacks against the Tor network.}
3. Recent works also experiment with techniques such as deep learning \cite{NASR2018} \cite{SIRINAM2018} to attack the Tor network. \ik{Deep learning and Artificial Intelligence are progressing rapidly and affecting other technologies on the way. The Tor research community must try to keep up with these new technologies as they can be powerful tools in the hands of real-life adversaries.}

\na{Our work provides an up-to-date review of the most important de-anonymisation attacks on Tor, providing insights in the evolution of these attacks}. 
We hope our review and the insights it provides will form a valuable resource to the wider Tor research community.

\section{Acknowledgements}
This work has been supported by the Cyber Security Research Centre Limited (CSCRC) whose activities are partially funded by the Australian Government's Cooperative Research Centres Programme.


\bibliographystyle{ieeetr}
\bibliography{main}

\end{document}